\documentclass[modern]{aastex631}
\usepackage{amsmath}

\received{June 23, 2024}
\revised{November 13, 2024}
\accepted{December 2, 2024}

\submitjournal{ApJ}

\shorttitle{SIDM, Core Collapse and the GGSL discrepancy}
\shortauthors{Dutra et al.}

\begin{document}

\title{Self-Interacting Dark Matter, Core Collapse and the Galaxy-Galaxy Strong Lensing discrepancy}

\author[0000-0001-7040-4930]{Isaque Dutra}
\affiliation{Department of Physics, Yale University, New Haven, CT 06511, USA}
\correspondingauthor{Isaque Dutra}
\email{isaque.dutra@yale.edu}

\author[0000-0002-5554-8896]{Priyamvada Natarajan}
\affiliation{Department of Astronomy, Yale University, New Haven, CT 06511, USA}
\affiliation{Department of Physics, Yale University, New Haven, CT 06511, USA}

\author[0000-0002-5116-7287]{Daniel Gilman}
\affiliation{Department of Astronomy $\&$ Astrophysics, University of Chicago, Chicago, IL 60637, USA}

\begin{abstract}
Gravitational lensing by galaxy clusters has emerged as a powerful tool to probe the standard Cold Dark Matter (CDM) paradigm of structure formation in the Universe. Despite the remarkable explanatory power of CDM on large scales, tensions with observations on small scales have been reported. Recent studies find that the observational cross-section for Galaxy-Galaxy Strong Lensing (GGSL) in clusters exceeds the CDM prediction by more than an order of magnitude, and persists even after rigorous examination of various possible systematics. We investigate the impact of modifying the internal structure of cluster dark matter sub-halos on GGSL and report that altering the inner density profile, given by $r^{\gamma}$, to steeper slopes with $\gamma > 2.5$ can alleviate the GGSL discrepancy. This is steeper than slopes obtained with the inclusion of the contribution of baryons to the inner regions of these sub-halos. Deviating from the $\gamma \sim 1.0$ cusps that CDM predicts, these steeper slopes could arise in models of self-interacting dark matter undergoing core collapse. Our results motivate additional study of sub-halo core collapse in dense cluster environments.
\end{abstract}



\section{Introduction} \label{sec:intro}

The concordance cosmological model of Cold Dark Matter (CDM) posits that dark matter behaves as a collision-less fluid, and forms gravitationally-bound halos with masses $M \sim 10^{14} M_{\odot}$, down to structures comparable to Earth mass $M \sim 10^{-6} M_{\odot}$. CDM halos form with an internal density distribution that is well fit by the Navarro-Frenk-White (NFW) profile \citep{Navarro+1997}.  Deviation of detected halo properties from these predictions could signal new physics, therefore many cosmic probes seek to characterize the properties of dark matter halos across a wide range of scales and cosmic environments.  

On large cosmological scales, observations agree remarkably well with CDM predictions \citep{Spergel+2007,PlanckCollaboration+2020-VI}. However, various tensions have emerged on sub-galactic scales, in dark matter halos below masses of $10^{11} M_{\odot}$ \citep[see][and references therein]{Bullock+2017}. On larger cluster-mass scales of $\sim 10^{14}M_{\odot}$, the ``arc statistics'' tension initially reported by \citep{Bartelmann+1998} highlighted the inability of CDM simulations to reproduce the abundance of observed lensed arcs, but several plausible resolutions to resolve this discrepancy have since emerged. Despite the variety of observables and cosmic environments associated with the small-scale and cluster-scale tensions, the common theme among them relates to an apparent failure of simulations to simultaneously reproduce the abundance and internal structure of dark matter halos predicted by CDM. Some of these tensions with CDM predictions were interpreted as a serious crisis for the theory and the underlying assumptions related to dark matter physics. In many if not most instances, however, the implications for dark matter physics became less clear upon further investigation of the implementation of baryonic feedback in galaxy formation simulations. These astrophysical processes can alter the properties of dark matter halos without requiring modifications to the fundamental nature of dark matter \footnote{see review Popolo \& Le Delliou 2017 \url{https://hal.archives-ouvertes.fr/hal-01324388v2/document} and references therein.}. 

In this paper, we review and revisit the report of an observational probe of small-scale structure, Galaxy-Galaxy Strong Lensing (GGSL), that reveals a brand-new tension with predictions of the CDM paradigm \citet{Meneghetti+2020} (hereafter M20). Comparing lensing observations with the Illustris-TNG cosmological simulation suite, we independently compute the GGSL probability adopting a different methodology from M20 and reaffirm its persistence. We then explore potential explanations to alleviate this discrepancy. 

The outline of our paper is as follows: in Section~2 we review the GGSL discrepancy; and outline the methodology including modeling of cluster lenses and the calculation of GGSL from lens mass models in Section~3. The properties of the observed lenses studied here are summarized in Section~4; and the computation of GGSL from the mass-matched simulated analogs from the Illustris-TNG suite is described in Section~5. Alteration of the inner density profile of cluster sub-halos explored with the Lenstronomy package is described in Section~6 and the Results of our analysis and our Conclusions \& Discussion are presented in Sections~7 \& 8 respectively.

\section{The Galaxy-Galaxy Strong Lensing Discrepancy}

Gravitational lensing by massive galaxy clusters offers a stringent test of the CDM paradigm of structure formation, both in terms of the number of collapsed halos -- the subhalo mass function -- and their density profiles (see the recent review by \citet{Natarajan+2024}). In this work, we focus on the relatively new tension with CDM in galaxy clusters first reported in M20. M20 showed that the measured GGSL cross section of galaxy clusters exceeds the value computed from hydrodynamical CDM simulations. For a fixed total number of cluster sub-halos, the GGSL cross section depends principally on the internal structure of the sub-halos. More centrally concentrated sub-halos are more likely to become super-critical and produce multiple images of  background sources. Previously, several works showed that observed clusters contain more high-circular-velocity sub-halos (i.e. sub-halos with maximum circular velocities $V_{\rm circ}>100$ km s$^{-1}$) compared to simulations \citep{Grillo+2015,Munari+2016,Bonamigo+2017,Meneghetti+2020}, although this feature was not quantified in terms of the strong-lensing cross section at that time. 

Mirroring the historical progression of apparent tensions with CDM on smaller scales, several studies highlighted possible solutions to the tension reported by M20 based on astrophysics or systematics in the simulations. These include numerical resolution effects \citep{Robertson2021} (see also \citet{vandenBosch+2018}), or choices related to the implementation of baryonic physics \citep{Bahe2021}. Further investigation found that neither of these factors resolves the GGSL tension without creating further new tensions between simulations and observations \citep{Ragagnin+2022,Meneghetti+2022,Tokayer+2024}. Resolutions proposed for the other small-scale tensions pertaining to the inner density profiles of low-mass galaxies through baryonic physics effects, in fact exacerbate the GGSL tension, which requires an enhancement, not a suppression, of the central density of cluster sub-halos.

Despite the significant advances of the past two decades in understanding the interplay between baryonic physics and dark matter physics inside halos\citep{DiCintio+2014a,DiCintio+2014b,Heinze+2023,Moline+2017,Moline+2022}, as well as the numerical advances in the modeling of these processes in clusters, at present no clear resolution to the GGSL discrepancy exists within the framework of CDM, as demonstrated in recent work by \citet{Tokayer+2024} and as explored here further.

The investigation by \citet{Tokayer+2024} examined two key issues: first, they altered the inner density of the pseudo-isothermal profile used to model the dark matter subhalo, which includes a core, produced by \texttt{Lenstool} and as employed in the GGSL computation for cluster lens models from M20. These modified profiles were adjusted to fit a truncated NFW (tNFW) profile, ensuring that they remained consistent with observed lensing constraints and adhered to the $\Lambda$CDM model. Second, they included the effects of the baryonic component and applied an extreme form of adiabatic contraction to the tNFW profiles\footnote{By using the publicly available code \texttt{Contra} found at \url{https://websites.umich.edu/~ognedin/contra/}}, as expected in $\Lambda$CDM. Including the modification due to the presence of baryons, produced a noticeable effect and \citet{Tokayer+2024} showed that it can lead to a steepening of the inner density slope of  utmost $\gamma < 2$ where $r^{\gamma}$ is the inner slope). This is consistent with the results found by \citet{Heinze+2023} when investigating baryonic effects on the inner density slope in the TNG50 simulation. However, even including these modifications, the computed GGSL probability is still significantly discrepant with observations and these modifications are insufficient to bridge the gap between simulations and lensing observations. In summary, even with adjustments to both the density profile and baryon-driven effects -- each designed to stay within the bounds of $\Lambda$CDM -- the calculated GGSL probability remained significantly discrepant, indicating the persistence of the tension between observations and simulations.

The tension reported by M20 implies that the concentration of matter in the inner regions of cluster sub-halos significantly exceeds the CDM prediction. As they note, this implies either missing physics in the simulations that would increase the cluster sub-halo lensing efficiency, or a modification to the internal structure of the sub-halos that raises their strong lensing cross section. In this paper, we investigate the GGSL under the assumption that modifications to the internal structure of cluster sub-halos are required to resolve the tension. 

\section{Methodology}

\subsection{Modeling Cluster Lenses}

We briefly outline the standard procedures used to model observed clusters-lenses using parametric mass profiles that are apt and amenable for direct and easy comparison with simulated mass-matched cluster analogs. More details of this procedure can be found in the review by \citet{Kneib+2011}.

\begin{itemize}
    \item The total mass distribution of a galaxy cluster is decomposed into parametric mass profile with components to represent larger scale and smaller scale contributions \citep{Natarajan+1997}:
    \begin{equation}\label{eqn:phi}
        \phi_{tot} = \sum_{i} \phi_i^{halo} + \sum_{j} \phi_j^{sub} + \phi_{\kappa,\gamma};
    \end{equation}
    where the first term represents large-scale smooth halos associated with the over-all cluster potential on scales of 100's of kpc; the second term represents smaller-scale perturbations on kpc scales that are associated with the subhalos of cluster member galaxies, and the last term is associated with a possible constant convergence or shear field, referred to as the mass-sheet degeneracy. Observed strong and weak lensing constraints including the positions, brightnesses and shapes of lensed background galaxies are used to constrain the overall mass distribution represented by Eqn.~\ref{eqn:phi}. It has been shown that measurements of the shear and convergence for extremely well calibrated lensing clusters with multiple families of strongly lensed images with spectroscopic redshifts such as the ones studied here, can be used to determine the mass sheet degeneracy term above \citep{Bradac+2004,Rexroth+2016,Bergamini+2021}. The components in Eqn.~\ref{eqn:phi} are determined using lens inversion algorithms, like for instance, the publicly available software package \texttt{Lenstool} that is widely used by the lensing community including by M20 and \texttt{lenstronomy}, a new package, that we use for the analysis presented here. Both the larger scale halos and subhalos are parameterized by self-similar \textit{dual Pseudo Isothermal Elliptical mass 
distributions} (dPIE profiles) in \texttt{Lenstool} \citep{Eliasdottir+2007}:
    \begin{equation}
        \rho_{dPIE}(r) = \frac{\rho_0}{(1+r^2/r_{\mathrm{core}}^2)(1+r^2/r_{\mathrm{cut}}^2)}
    \end{equation}
    
    \item The enclosed projected total mass can be found with
    \begin{equation}
        \mathrm{M}_{\mathrm{2D}}(R) = 2 \pi \Sigma_0 \frac{r_{\mathrm{core}} r_{\mathrm{cut}}}{r_{\mathrm{cut}}-r_{\mathrm{core}}} \left( \sqrt{r_{\mathrm{core}}^2+R^2} - r_{\mathrm{core}} - \sqrt{r_{\mathrm{cut}}^2+R^2} + r_{\mathrm{cut}} \right)
    \end{equation}
    
    \item where the projected surface mass density $\Sigma(R)$ is given by:
    \begin{equation}
        \Sigma(R) = \Sigma_0 \frac{r_{\mathrm{core}} r_{\mathrm{cut}}}{r_{\mathrm{cut}}-r_{\mathrm{core}}} \left( \frac{1}{\sqrt{r_{\mathrm{core}}^2+R^2}} - \frac{1}{\sqrt{r_{\mathrm{cut}}^2+R^2}} \right),
    \end{equation}
    \item $\rho_0$ and $\Sigma_0$ can be related to the fiducial velocity dispersion from
    \begin{align}
        \sigma^2_{dPIE} &= \frac{4}{3} G \pi \rho_0 \frac{r_{\mathrm{core}}^2 r_{\mathrm{cut}}^3}{(r_{\mathrm{cut}}-r_{\mathrm{core}})(r_{\mathrm{cut}}+r_{\mathrm{core}})^2} \\
        &= \frac{4}{3} G \Sigma_0 \frac{r_{\mathrm{core}} r_{\mathrm{cut}}^2}{r_{\mathrm{cut}}^2-r_{\mathrm{core}}^2}
     \end{align}

    where $\sigma_{dPIE}$ is found from the lens model optimization obtained by reproducing observed lensed image constraints and is related to the measured central velocity dispersion of cluster galaxies via $\sigma^2_{dPIE} = \frac{2}{3} \sigma^2_0$.    
    The following substitution is done to generalize to the elliptical case, $R \rightarrow \tilde{R}$, profiles that are available for use within \texttt{Lenstool}:
    \begin{equation}
        \tilde{R}^2 = \frac{X^2}{(1+\epsilon)^2} + \frac{Y^2}{(1-\epsilon)^2}
    \end{equation}
    where the ellipticity $\epsilon = (A-B)/(A+B)$, where A and B are the semi-major and semi-minor axis respectively. The ellipticity as defined in \texttt{Lenstool}, has a slightly different definition as $\hat{\epsilon}$ that is given by:
    \begin{equation}
        \hat{\epsilon} = \frac{A^2 - B^2}{A^2+B^2}.
    \end{equation}
    The dPIE profiles in \texttt{Lenstool} are parameterized by projected sky positions $x$ and $y$, fiducial velocity dispersion $\sigma_{dPIE}$, ellipticity $\hat{\epsilon}$, scale radius $r_{\mathrm{core}}$, cut radius $r_{\mathrm{cut}}$, and orientation $\theta_{\hat{\epsilon}}$. The best-fit cluster lensing model provides constraints on the total mass enclosed within $r_{\mathrm{cut}}$ of the larger scale halos as well as the  integrated mass within $r_{\mathrm{cut}}$ for the subhalos. How the mass and surface mass density are distributed radially within subhalos is not a quantity that can be currently constrained with the data in hand. Furthermore, to even obtain this integrated constraint on the mass enclosed within the aperture for subhalos, we need to collapse the number of parameters required to describe galaxy-scale subhalos and the associated cluster member galaxies. To do this, typically,  an explicit relation between mass and light that is empirically well motivated for cluster galaxies is assumed. The following scaling relations with luminosity $L$ are adopted involving the characteristic subhalo parameters ($\sigma_{\mathrm{dPIE}}$, $r_{\mathrm{cut}}$) \citep[see]{Eliasdottir+2007}:
    \begin{align}
        \sigma_{dPIE} &= \sigma_{dPIE}^{ref} \left( \frac{L}{L_{*}} \right)^\alpha \\
        r_{cut} &= r_{cut}^{ref} \left( \frac{L}{L_{*}} \right)^\beta.
    \end{align}    
\end{itemize}
Therefore, for a well constrained lens model, masses are obtained for all constituent subhalos using the above calibrated scaling relation; with these, the overall subhalo mass function can be determined for a given cluster. As noted in previous works, such as \citet{Natarajan+2017}, the lensing determined subhalo mass function for clusters is in very good agreement with the one computed from mass-matched simulated analogs, as shown in Fig.~\ref{fig:shmf_all}.

\subsection{Calculation of the GGSL} \label{sec:calculation_ggsl}

In this section, we briefly define the GGSL probability and outline its calculation. Starting from the surface mass density maps of the lensing clusters obtained as noted above, we are interested in computing the probability that a background source could be strongly lensed by foreground substructures, namely, subhalos in the cluster.

\begin{figure}[ht]
    \centering
    \includegraphics[width=1.0\textwidth]{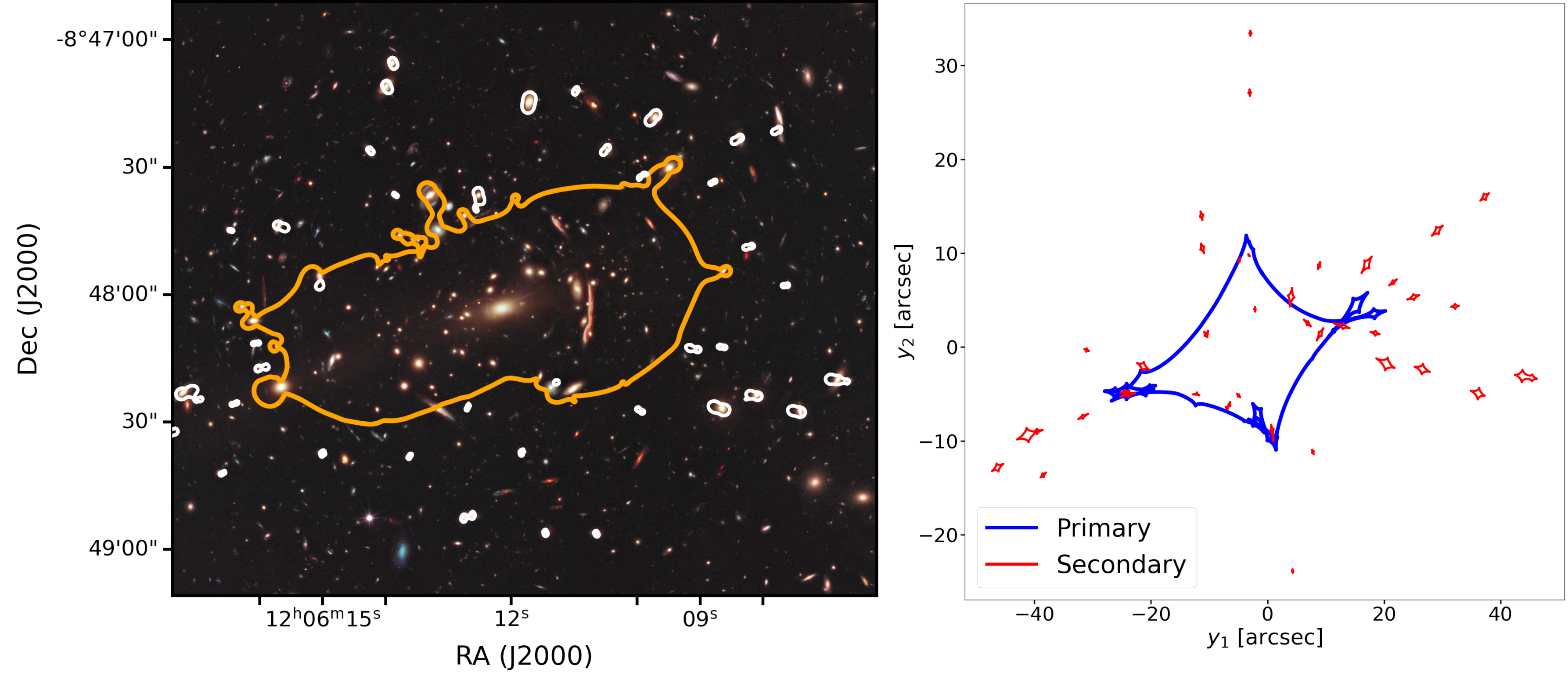}
    \caption{The critical lines and caustics for the cluster MACS J1206: calculated using \texttt{lenstronomy} these are shown in the left and right panel, respectively. The primary critical line is identified in orange, while secondaries are shown in white; similarly, primary caustics are identified in blue, while secondary caustics are shown in red. Image credit: The color composite of MACS J1206, NASA, ESA, M. Postman (STScI) and the CLASH Team. We note here that the field of view (FOV) in the left panel is not the same as the FOV in the right panel. The caustics are smaller and harder to see, therefore, we present a zoomed in view in the right panel for visibility.}
    \label{fig:m1206_cc}
\end{figure}

All lensing quantities of interest can be derived from the projected Newtonian potential $\phi$ related to the projected surface mass density $\Sigma$ as $4\pi G \Sigma = \nabla^2 \phi$. These can be directly obtained from the derived lensing mass maps. The deflection angle $\alpha$ is found from the partial derivative of the projected potential while computed second order derivatives characterize induced deformations: the convergence $\kappa$ quantifies the isotropic deformation and affects the apparent size and brightness of the lensed objects, while the shear $\gamma$ describes the anisotropic deformation affecting the stretching along the shear direction. The convergence $\kappa$ and shear $\gamma$ define the magnification $\mu^{-1} = (1-\kappa)^2 - \gamma^2 = (1-\kappa-\gamma)(1-\kappa+\gamma)$, which is infinite when $\kappa \pm \gamma = 1$. The set of points of infinite magnification in the image plane define two closed lines, the \textit{critical lines}, with \textit{caustics} being the corresponding lines projected onto the source plane. For a simple mass distribution, the inner critical line can be identified as the \textit{radial} critical line, where background sources are deformed radially, and the outer one as the \textit{tangential} critical line, where background sources are tangentially deformed into arcs and rings. Since the radial critical lines of massive cluster members are expected to be insignificant in terms of area enclosed within \cite[see][Supplementary materials]{Meneghetti+2020}, we consider only the \textit{tangential critical lines} where $\kappa + \gamma = 1$. We plot the tangential critical lines and caustics for one of the clusters studied here, MACS J1206, in Fig.~\ref{fig:m1206_cc}. 

Critical lines can be assigned an Einstein radius $\theta_\mathrm{E}$ given their average enclosed area $\mathrm{A}$, $\theta_\mathrm{E} = \sqrt{\mathrm{A}/\pi}$. Following the same classification criteria adopted in \citet{Meneghetti+2022}, we identify critical lines with $\theta_\mathrm{E} > 5"$ as \textit{primary}; those with $0.5" < \theta_\mathrm{E} < 3"$ as \textit{secondary}, and exclude the rest in our analysis. We adopt the same definitions as M20 for consistency for comparison. These selection criteria permit us to filter subhalos with masses in the range of $\sim 10^{10-12}h^{-1}\,\mathrm{M}_\odot$. And by definition, GGSL events originate from the secondary critical lines associated with cluster subhalos in this mass range. Substructure associated with Einstein radii less than $0.5"$ cannot be resolved by simulations, while those with Einstein radii ranging between $3"$ and $5"$ are associated with larger, group-scale masses  and not individual galaxies \citep{Meneghetti+2022,Ragagnin+2022,Meneghetti+2023}.
The GGSL cross section $\sigma_\mathrm{GGSL}$ is defined as the total area enclosed by all secondary caustics. Since both the convergence $\kappa$ and shear $\gamma$ depend on the source redshift $z$, $\sigma_\mathrm{GGSL} = \sigma_\mathrm{GGSL}(z)$ is also a function of redshift. The probability of GGSL events $P_\mathrm{GGSL}$ is defined as the GGSL cross section normalized by the Field of View (FOV) area on the source plane $\mathrm{A}_\mathrm{s}$, $P_\mathrm{GGSL} = \sigma_\mathrm{GGSL}/\mathrm{A}_\mathrm{s}$.

In summary, the caustics associated with subhalos mark regions where background galaxies will be strongly lensed and dividing the caustic area by the FOV source plane area defines the probability of GGSL events. It is this derived probability that M20 found to be higher by more than an order of magnitude for observed lensing clusters compared to their simulated counter-parts in $\Lambda$CDM.

Here, we adopt a new approach and utilize the recently developed open source gravitational lensing python package \texttt{lenstronomy} \citep{Birrer+2018,Birrer+2021} to perform the lensing calculations. For each cluster, we first load the list of best-fit potentials and their optimized characteristic properties obtained  from \texttt{Lenstool} representing all cluster members and initialize the appropriate analytical potentials in \texttt{lenstronomy}. For a given source redshift, the deflection map and its relevant derivatives are calculated as are the maps for the convergence $\kappa$ and shear $\gamma$. The contours where $\kappa + \gamma = 1$ are found utilizing the open source package \texttt{scikit-image}\citep{VanDerWalt+2014}, and by projecting them onto the source plane, the associated caustics are found. Consistent with M20, we delimit the FOV to $200"\times200"$ for all the clusters studied here. Then the areas associated with secondary caustics are summed and normalized by the projected FOV area to compute the $P_\mathrm{GGSL}(z_s)$ for source redshifts $z_s$ between 0.5 and 7.

We calculate the properties of cluster sub-halos required to reproduce the observed GGSL cross section. This work expands on the earlier analysis by \citet{Yang+2021}, who showed that core collapse in Self-Interacting Dark Matter (SIDM) models raises the central density of cluster sub-halos and can hence increase the GGSL cross section. Here, we explore more comprehensively the consequence of core collapse by calculating the strong-lensing cross section with the convergence maps of simulated galaxy clusters. To examine the degree to which feedback processes can impact the computed GGSL, we analyze a new simulation suite - the Illustris-TNG suite \footnote{More information on the TNG suite can be found at \url{https://www.tng-project.org/}} - for comparison with lensing data that implements feedback differently from the DIANOGA suite used for comparison by M20. Unlike in DIANOGA, in the Illustris suite of simulations, feedback from star formation and accreting black holes is modeled very differently. Stellar feedback is driven by galactic-scale outflows and multi-modal feedback from accreting black holes that operate in a thermal quasar mode at high accretion rates and a kinetic wind mode at low accretion rates inform their sub-grid feedback physics models. These implementations of feedback are calibrated to quench gas cooling and star formation that is required to match the properties of observed galaxies. These astrophysical processes are relevant to GGSL as they potentially serve to rearrange mass from the inner regions of halos.

In what follows, we compute the GGSL cross section for sample of simulated clusters that are matched in total mass enclosed within $R_{200}$ \footnote{where $R_{200}$ is the radius within which the density is 200 times the critical density of the Universe at the halo's redshift $z$} to a set of observed lenses (see Table~\ref{tab:sample} for the list). We calculate the GGSL from simulated clusters using the software package \textit{lenstronomy} \footnote{\url{https://github.com/lenstronomy/lenstronomy}} \citep{Birrer+2018,Birrer+2021}, a different code than the {\tt{Lenstool}} software used by M20\footnote{\url{https://projets.lam.fr/projects/lenstool/wiki}}. Using {\tt{lenstronomy}}, we modify the properties of Illustris-TNG CDM cluster sub-halo density profiles to more centrally-concentrated density profiles. In particular, we consider a density profile given by \citet{Munoz+2001}:
\begin{equation}
\label{eqn:gnfw_density_profile}
    \rho_\mathrm{gNFW}(r) = \frac{\rho_0}{(r/r_\mathrm{s})^\gamma (1+(r/r_\mathrm{s})^2)^{(\beta-\gamma)/2}}
\end{equation}
(hereafter the generalized NFW, or gNFW, profile) where $\gamma$ and $\beta$ are the logarithmic inner and outer slopes, respectively. An NFW profile in CDM corresponds to  $\gamma=1$ and $\beta = 3$.\footnote{The density profile in Eqn.~\ref{eqn:gnfw_density_profile} differs from an NFW profile through the $\left(1+(r/r_\mathrm{s})^2\right)$ in the denominator, whereas the NFW profile has $\left(1+(r/r_\mathrm{s})\right)^2$. This detail however is not relevant to our main results.} 

To derive the values of $\gamma$ and $\beta$ in the density profile given in Eqn.~\ref{eqn:gnfw_density_profile} that are required to match the observed GGSL, we take 3 separate projections of a simulated galaxy cluster analog and replace cluster sub-halos with objects possessing gNFW density profiles. This is implemented in such a way as to conserve the total mass of the cluster, while re-distributing the mass inside the sub-halo. We then compute the GGSL probability with these steepened sub-halo profiles.

In addition to calculation of the GGSL outlined above that was performed as done previously by M20 for comparison, in this work we compute the GGSL probability directly from the particle data in the TNG-300 simulation. First we compute using dark matter particles that include the adiabatic contraction from the baryonic component and then we explicitly include the dark matter and baryonic particles to compute the GGSL. We find very good agreement with the prior calculation done with NFW profile fits for the simulated clusters. Further details are provided in the Appendix in Section \ref{sec:ggsl_particle}.

\section{Properties of the observed cluster lens sample}\label{sec:ggsl_cluster}

With the intention of independently verifying the GGSL probability discrepancy reported by M20, we select five observed cluster-lenses to study as our sample. Four of them have well-constrained mass distributions (lensing mass models) derived from the Hubble Frontier Fields (HFF) program data. These four were also selected by M20, and are: Abell 2744 (A2744) at $z=0.308$ \citep{Merten+2011,babyk+2012,Jauzac+2016}; Abell S1063 (AS1063) at $z=0.348$ \citep{Gomez+2012,Gruen+2013}; MACS J1206.2-0847 (M1206) at $z=0.439$ \citep{Umetsu+2012,Biviano+2013} and MACS J0416.1-2403 (M0416) at $z=0.397$ \citep{Jauzac+2014}.

The mass models for these four HFF clusters computed using \texttt{Lenstool} are publicly available on the MAST Archive \footnote{\url{http://archive.stsci.edu}}. Several groups using independently developed lens inversion algorithms provided magnification and mass maps for the HFF clusters. For our work here, we used the mass models provided by the CATS collaboration. Further details of the \texttt{Lenstool} reconstructions from the CATS collaboration can be found in several papers including but not limited to \citet{Jauzac+2015,Jauzac+2016,Mahler+2018,Bergamini+2019,Bergamini+2021,Meneghetti+2022}. The fifth cluster in our sample, PSZ1 G311.65-18.48 (PG311), also included as part of the sample in \citet{Meneghetti+2022}, is a massive cluster that was selected differently. It was selected  not for its lensing properties, but from the ESA Planck catalog of clusters that produce a detectable Sunyaev-Zeldovich (SZ) decrement signal \citep{PlanckCollaboration+2014-XXIX}. The properties of our selected cluster sample and their properties are summarized in Table~\ref{tab:sample}.

\begin{table}[ht]
\begin{tabular}{cccc}
\hline
Cluster     & Redshift & $\mathrm{M}_{200}$ Estimate [$10^{15} M_\odot$] & Subhalos  \\ \hline
Abell 2744  & 0.308    & $2.22^{+0.13}_{-0.12}$                          & 246        \\
Abell S1063 & 0.348    & $3.97^{+1.6}_{-0.9}$                            & 222        \\
MACS J0416  & 0.397    & $1.60 \pm 0.01$                                 & 191        \\
MACS J1206  & 0.439    & $1.4 \pm 0.2$                                   & 258        \\
PSZ1 G311   & 0.444    & 0.85                                            & 194        \\ \hline
\end{tabular}
\caption{Summary of the observed cluster lenses in our sample.}
\label{tab:sample}
\end{table} 

{\bf Abell 2744} (also known as AC118, or RXCJ0014.3-3022) at $z \sim 0.308$, is an exceedingly complex system undergoing a complicated merging event, and has been dubbed Pandora's Cluster. It displays a total of about 34 strongly lensed images identified in 11 multiple-image systems \citep{Merten+2011} and was chosen for the HFF by \citep{Lotz+2017} and more recently, also observed as part of the JWST UNCOVER program \citep{Bezanson+2022}. The total enclosed mass within a radius of 1.3 Mpc for Abell 2744 is measured to be $(1.8 \pm 0.4) \times 10^{15} M_\odot$ by \citet{Merten+2011} pre-HFF; and with HFF data \citet{Jauzac+2015} measure the total cluster mass within 250 kpc as $(2.765 \pm 0.008) \times 10^{14} M_\odot$ and enclosed within 1.3 Mpc as $(2.3 \pm 0.1) \times 10^{15} M_\odot$ using strong and weak lensing analysis in \citet{Jauzac+2016}. Here, we use the mass estimate of $M_{200} = 2.22^{+0.13}_{-0.12} \times 10^{15} M_\odot$ within $R_{200} = 2.38^{+0.36}_{-0.31} \mathrm{Mpc}$. The best-fit lensing model for this cluster  comprises 258 total mass components, of which six are large scale mass concentrations; six are large scale potentials that lie outside the Hubble Advanced Camera for Surveys FOV but are in the vicinity and need to be included and the rest correspond to confirmed cluster member galaxies.
 
{\bf Abell S1063} at $z \sim 0.348$, is a particularly interesting cluster due to its unusually high X-ray luminosity that indicates that it recently underwent a major merging event \citep{Gomez+2012}. Part of the HFF, it was also observed previously by HST in the Cluster Lensing And Supernova survey with Hubble (CLASH) program \citep{Postman+2012}, and spectroscopically followed up by the  CLASH-VLT program \citep{Caminha+2016}. The estimated mass is $M_{200} = 3.31^{+0.96}_{-0.68} \times 10^{15}M_\odot$ \citep{Williamson+2011,Gomez+2012,Gruen+2013,Diego+2016}. The best-fit lensing mass model for Abell S1063 comprises 227 total mass components, of which two represent large scale central mass concentrations and the rest are all associated with confirmed cluster member galaxies.

{\bf MACS J0416.1-2403} at $z \sim 0.397$ is also a massive cluster in a complex dynamical state: it features a bimodal velocity distribution, showing two central peaks and elongated sub-clusters suggesting that they are being observed in a pre-collisional but strongly interacting phase \citep{Balestra+2016}. It was also imaged by the HFF and CLASH programs, and followed up as part of the CLASH-VLT program. Its total projected mass inside a 200 kpc aperture was estimated to be $(1.60 \pm 0.01) \times 10^{14} M_\odot$ by \citet{Jauzac+2014}, while the CLASH analysis by \citet{Umetsu+2016} estimates $M_{200} = (1.412 \pm 0.366) \times 10^{15}$, and \citet{Balestra+2016} report $M_{200} = (1.12 \pm 0.26) \times 10^{15} M_\odot$ using CLASH-VLT. The best-fit lensing mass model for this cluster contains 197 total mass components, of which six are large scale central mass concentrations and the rest are associated with confirmed cluster member galaxies.

{\bf MACS J1206.2-0847} at $ z \sim 0.440$ was previously studied by \citet{Ebeling+2009}. The CLASH analysis by \citet{Umetsu+2016} estimates the cluster mass to be $M_{200} = (2.596 \pm 0.604) \times 10^{15} M_\odot$. The obtained best-fit lensing mass model contains 264 total mass components, of which three are large scale central potentials; an additional component of external shear and the rest are associated with confirmed cluster member galaxies.

The final selected cluster in our sample is {\bf PSZ1 G311.65-18.48} (PSZ1 G311), $z \sim 0.443$. Our current sample has some overlap with the sample analyzed in M20, but also includes PSZ1 G311, which is not selected via strong-lensing as noted above, but detected in the ESA Planck catalog of Sunyaev-Zeldovich sources \citep{PlanckCollaboration+2014-XXIX}. It features a remarkable tangential arc, the highly distorted image of a distant star-forming galaxy at $z \sim 2.370$ lensed 12 times, dubbed the \textit{Sunburst Arc} by \citet{Rivera-Thorsen+2017}. \citet{Sharon+2022} estimate the projected cluster mass within 250 kpc to be $2.93^{+0.01}_{-0.02} \times 10^{14} M_\odot$ using HST imaging, archival VLT/MUSE spectroscopy, and Chandra X-ray data, while \citet{Dahle+2016} estimate its mass using Planck SZ data to be $M_{500} = 6.6^{+0.9}_{-1.0} \times 10^{14} M\odot$. Converting to $M_{200}$ using a typical concentration-mass relation, \citet{Meneghetti+2022} estimated $M_{200} \sim 8.5 \times 10^{14} M_\odot$. The best-fit lensing mass model contains 202 total mass components, of which eight are large scale central potentials and the rest are associated with confirmed cluster member galaxies.

We note that our selected sample of clusters is very similar to the one selected by \citet{Meneghetti+2022} and identical to the clusters selected by \citet{Tokayer+2024}. However, using the best-fit list of mass components from the optimized lensing mass model, here we use both \texttt{Lenstool} and \texttt{lenstronomy} to calculate the GGSL probability for each of these clusters, as outlined in Section~\ref{sec:calculation_ggsl}. Our results, computed independently and in a distinct way are shown in Fig.~\ref{fig:ggsl_sidm}. Our results are in excellent agreement with the results reported in M20.


\section{Computing GGSL from mass-matched simulated analogs in Illustris-TNG}\label{sec:ggsl_cdm}

We calculate the GGSL probability of subhalos in cluster-scale halos in the Illustris-TNG simulations by modeling them with analytical mass density profiles. The Illustris-TNG simulations are a suite of cosmological magnetohydrodynamical simulations of galaxy formation \citep{Nelson+2019}\footnote{Publicly available at \url{https://www.tng-project.org}}. In this study, we focus on TNG300, with a periodic box with side length of 302.6 Mpc, particle/cell number of $2500^3$ for baryons and dark matter; baryonic particle mass of $1.1 \times 10^7 M_\odot$; and dark matter particle mass of $5.9 \times 10^7 M_\odot$.

For each cluster in our observational sample, we identify three best total mass-matched halo analogs in TNG300 with comparable $M_{200}$'s at the same redshift as the observed sample. For the TNG300 cluster halos, $M_{200}$ is estimated using the friends-of-friends FoF algorithm. Our simulated analogs have a mean mass $\sim 10^{15} M_\odot$.

We model the central halo and subhalos of the selected TNG300 analogs with the NFW profile, which has been shown to be a good fit for DM halos over a wide range of halo masses \citep{Navarro+1997}. It is particularly useful since the profile can be specified with two parameters, which can be chosen to be the radius $R_{200}$ which encloses an average density of $200 \rho_\mathrm{c}$ and the scale radius $R_\mathrm{s}$, where the profile's logarithmic slope steepens from $-1$ to $-3$:

The density profile is given by:
\begin{equation}
    \rho_\mathrm{NFW}(r) = \frac{\delta_\mathrm{c} \rho_\mathrm{c}}{(r/r_\mathrm{s})(1+r/r_\mathrm{s})^2},
\end{equation}
where $\rho_\mathrm{c} \equiv 3 H^2(z)/(8\pi G)$ is the critical density of the Universe at the halo's redshift $z$, and $\delta_\mathrm{c}$ is the characteristic over-density of the halo, defined as:

\begin{equation}
    \delta_\mathrm{c} \equiv \frac{200}{3} \frac{c^3}{\ln(1+c) - c/(1+c)}
\end{equation}
 $c$ is the characteristic concentration of the halo, defined as $c \equiv R_{200}/R_\mathrm{s}$.

We adopt the same procedure to calculate the GGSL probability for the simulated analogs as done for observed clusters described in Section~\ref{sec:ggsl_cluster}. For each subhalo belonging to the same TNG300 Group, we now initialize an NFW profile in \texttt{lenstronomy} with the halo's parameters $M_{\mathrm{200}}$ and concentration $c$, wherein the scale radius $R_\mathrm{s}$ is found from the relationship $R_\mathrm{max} \simeq 2.163 R_\mathrm{s}$ \citep[see Sec 11.1.2]{vanDenBosch+2010-book}, and $R_\mathrm{max}$ is the radius at which the halo's particles attain maximum velocity.

 Furthermore, we project each simulated cluster analog and its subhalos along three independent lines-of-sight corresponding to the simulation's three spatially orthogonal planes to average over. We assume $M_{\mathrm{200}} \approx M_\mathrm{sub}$, the total mass of all bound member particles. We include only subhalos with virial mass $M_{\mathrm{200}}$ between $10^{10} h^{-1} M_\odot$ and $10^{12} h^{-1} M_\odot$ as done in M20, in addition to the massive central large scale halos. The GGSL probability of the TNG300 cluster analogs is then  calculated as described in Section~\ref{sec:calculation_ggsl} and the results are shown in Fig.~\ref{fig:ggsl_sidm}. As seen, our results confirm the reported tension with CDM clearly demonstrating the order of magnitude gap.

As noted above, we also compute the GGSL for the simulated analogs to our observational sample using the particle data (rather than the halo catalogs) directly and find that the GGSL discrepancy remains. For each matched simulated analog, we projected the smoothed particle data as described in Section \ref{sec:ggsl_particle} and computed the GGSL probability for 3 distinct cases: i) including only the dark matter particles from the TNG300-Dark-1 simulation, ii) Dark matter only particles from the TNG300-1 simulation that includes baryons, in an effort to capture the effect of baryonic-induced adiabatic contraction of the dark matter halos, iii) explicitly including both the dark matter and baryonic particles from the TNG300-1 simulation. The resulting GGSL probabilities are shown in Fig. \ref{fig:ggsl_tng_particle}. The computed caustic maps for the above cases are provided in Section \ref{sec:ggsl_particle}. Furthermore, we also examined the three orthogonal projections for each analog, including the projection along the direction of elongation, and find the difference in GGSL probabilities to be negligible.


\section{Altering inner density profiles to alleviate the GGSL discrepancy}

While baryonic physics can, in principle, impact the central density profiles of halos \citep{DiCintio+2014a,DiCintio+2014b,Heinze+2023,Moline+2017,Moline+2022}, the structure of a dark matter halo can evolve more dramatically if one relaxes the assumption of collision-less dynamics inherent to CDM. The class of theories with this attribute is commonly referred to as self-interacting dark matter (SIDM). Dark matter self-interactions were initially proposed as a mechanism to generate flat cores inside halos to explain the low central densities (relative to CDM predictions) of low-mass dwarf galaxies \citep{Spergel+2000}. Soon after the initial SIDM proposal, \citet{Balberg+2002} pointed out that these dark self-interactions could also result in an opposite effect, due to the `gravothermal catastrophe', a phenomenon initially studied by \citet{Lynden-Bell+1968} in the context of dense star clusters. The gravothermal catastrophe, or `core collapse' as it is now more commonly phrased, refers to a runaway contraction of halo density profiles. Core collapse represents the ultimate fate of all halos in SIDM unless an external heat reservoir supplies the central core with sufficient energy to remain intact \citep[e.g.][]{Zeng+2022,Slone+2023}. SIDM models have gained traction on small scales because the processes of core formation and eventual core collapse give rise to a diversity of density profile slopes in sub-halos that appears more consistent with the properties of low-mass galaxies \citep{Zavala+2019,Sales+2022,Yang+2023}. 

The process of core collapse is particularly relevant for lensing because core collapse raises the central density, and therefore the lensing efficiency, of a halo. Recently, \citet{Minor+2021} reported the detection of dark object with a central density orders of magnitude higher than the CDM expectation, which could result from core collapse in an SIDM halo \citep{Nadler+2023}. Core collapse can also produce a distinct signal in the magnifications among images of quadruply-imaged quasars \citep{Gilman+2021,Gilman+2023}. These studies focus on strong lensing by isolated individual galaxy-scale lenses, unlike the case of GGSL in clusters. Cosmological simulations of SIDM on cluster scales that permit direct comparison and calculation of the GGSL probability are currently unavailable as there are multiple numerical challenges with resolution and convergence that are yet to be surmounted \citep[e.g.][]{Fischer+2024,Mace+2024,Ragagnin+2024}. 

\subsection{Mimicking core-collapse within lenstronomy}\label{sec:ggsl_sidm}

It is clear that more centrally concentrated subhalos likely enhance GGSL and could therefore alleviate the GGSL discrepancy. This however requires steepening the inner density profiles in subhalos beyond $\rho \propto r^{-1}$ of the NFW profile characteristic of CDM. Cluster galaxies are observed to have a strong concentration of baryons in the inner regions, however as recently shown by \citet{Tokayer+2024}, the modification to the inner density profiles from these effects cannot account for the GGSL discrepancy. In \citet{Tokayer+2024}, the density profiles of cluster galaxies were adjusted to account for extreme baryon-driven adiabatic contraction, while still remaining consistent with $\Lambda$CDM. However, the resulting change in computed GGSL probability was insufficient to resolve the observed discrepancy.

Such additional steepening is expected to occur in beyond-CDM dark matter models that contain self-interactions, SIDM models. The class of SIDM models in which appreciable steepening occurs are ones in which gravothermal core collapse has commenced in the inner-most regions.  In order to mimic this process of core collapse that results in a more centrally concentrated halo, we utilize the generalized NFW (gNFW) profile and alter the inner and outer slopes of the mass density profile for simulated subhalos in the simulated analogs. The larger scale central, smooth halo mass density profiles are kept unaltered and are retained with an NFW fit as they account for the well-constrained large scale lensing features in the observed clusters. 

We choose the gNFW profile \citep{Munoz+2001} that is implemented in \texttt{lenstronomy} to steepen inner density profiles:

\begin{equation}
    \rho_\mathrm{gNFW}(r) = \frac{\rho_0}{(r/r_\mathrm{s})^\gamma (1+(r/r_\mathrm{s})^2)^{(\beta-\gamma)/2}}
\end{equation}
where $\gamma$ and $\beta$ are the logarithmic inner and outer slope, respectively. Compared to standard cusped models such as the NFW profile where 3D density profile $\rho \propto (r+r_\mathrm{s})^{(\beta-\gamma)}$, the gNFW profile implemented in \texttt{lenstronomy} follows $\rho \propto (r^2+r_\mathrm{s}^2)^{(\beta-\gamma)/2}$ and possesses convenient analytic properties that facilitate lensing calculations \citep{Munoz+2001}.

After loading all TNG300 subhalos into \texttt{lenstronomy} as described in the previous section for each observed cluster analog, we modify the mass density profile from the CDM NFW profile into a generalized NFW (gNFW) profile for the subhalo population while conserving mass within $R_{200}$. This \texttt{lenstronomy} profile conversion procedure is done through the open source python package \texttt{pyHalo}\footnote{Publicly available at \url{https://github.com/dangilman/pyHalo}}. We start by calibrating the gNFW profile by setting $\gamma=1$ and finding the logarithmic outer slope $\beta$ such that the caustic area of the subhalo matches the caustic area produced by an NFW profile; this we find is the case when $\beta=2.84$. The outer slope is now fixed to this value and then we alter the inner slope $\gamma$ which mimics various stages of core collapse in generic SIDM models. Next, we recalculate the GGSL probability for each simulated cluster analog by fitting a gNFW profile varying $\gamma=1.0; 2.0; 2.5; 2.9; 2.99$ for the entire subhalo population as a whole. All the subhalos in the simulated analog cluster are assigned the same steepened inner slope. The core-collapse cases with $\gamma = 2.5; 2.9$ are shown in Fig.~\ref{fig:ggsl_sidm}. 

The difference in the computed critical lines of a TNG300 cluster analog fitted with an NFW profile and the gNFW with $\gamma=2.9$ can be seen in Fig.~\ref{fig:tng_critical_lines}. As seen, the secondary critical lines for the gNFW profiles representing the core-collapsed subhalos are significantly larger and better resemble the critical lines of observed clusters (see for example the case of MACS J1206 shown in Fig.~\ref{fig:m1206_cc}). We also calculate the GGSL probability for the final stage of gravothermal core collapse to a point mass as this represents the limiting case. In \texttt{lenstronomy}, this case is initialized by specifying the Einstein radius $\theta_\mathrm{E}$ of the point mass with mass $M$,

\begin{equation}
\theta_\mathrm{E}^2 = \frac{4 G M}{c^2} \frac{D_\mathrm{LS}}{D_\mathrm{S} D_\mathrm{L}}
\end{equation}
where $D_\mathrm{LS}$ is the angular diameter distance between the lens and source plane, $D_\mathrm{L}$ between the observer and the lens plane, and $D_\mathrm{S}$ between the observer and the source plane. The GGSL probability of the core-collapsed cases c1-SIDM and c2-SIDM and that of the point mass are plotted in Fig.~\ref{fig:ggsl_sidm}.
For reference, the critical lines and caustics computed from the projected smoothed particle data of the same analogs are shown in Fig. \ref{fig:tng_particle_caustics}.

\section{Results}

We compare lensing observations with a new set of cosmological simulations, TNG300 from the Illustris suite and compute GGSL using a new modeling procedure, distinct from M20, to verify that the tension persists independently of simulations and modeling assumptions. We find that the GGSL discrepancy persists for cluster lenses, even with our use of a new sample of mass-matched simulated clusters from the TNG300 simulation suite and computation performed using an independent method utilizing the {\tt{lenstronomy}} software package instead of {\tt{Lenstool}}, as done previously by M20. Our observed cluster sample has common candidates with the M20 sample, but also includes a massive cluster PSZ1 G311.65-18.48 that has been selected not for its observed lensing properties, but rather from the Sunyaev-Zeldovich decrement it produced as detected by the Planck satellite \citep{PlanckCollaboration+2014-XXIX}. With this new methodology and use of an entirely independently generated Illustris simulation suite for comparison, the computed GGSL still remains discrepant by an order of magnitude, as reported previously. We also quantify the impact of projection effects and the inclusion of sub-halo ellipticity on the computed GGSL for simulated sub-halos, and find no significant increase in the GGSL probability from either factor. Therefore, projection effects and the inclusion of ellipticity for cluster sub-halos cannot bridge the order-of-magnitude GGSL discrepancy.

We now investigate how modifying the central density profile of cluster sub-halos affects the GGSL cross section. We note that explicitly including the contribution of stars in the inner regions also alters CDM density profiles, but this effect results in $\gamma < 2.0$, insufficient to completely alleviate the GGSL discrepancy \citep{Tokayer+2024,Heinze+2023}.

The modifications we make to re-arrange the mass in the inner regions of simulated sub-halo density profiles are intended to replicate the various stages of on-going core-collapse predicted by SIDM theories. We vary the inner logarithmic slopes $\gamma$ in the profile shown in Eqn.~1, starting initially with the CDM NFW inner slope of $1$, and steepening it to $3$. Simulations of SIDM sub-halos exhibit a variety of inner profile slopes that fall within this range \citep{Nadler+2023}. 

We find that the GGSL probability increases as the inner slope steepens, as noted by \citet{Yang+2021}. The median GGSL probability depends on the inner slope $\gamma$: when $\gamma = 1.5$ the GGSL probability is roughly 1.5 times higher than for CDM; for $\gamma=2$ it increases to roughly 2 times that of CDM; for $\gamma=2.5$ it is 3.5 times and for the near ultimate stage of core collapse with $\gamma=2.9$ it is found to increase by a factor of 8, bringing the GGSL cross section from simulations into agreement with what is found for observed cluster lenses. We draw attention to the fact that the radial distribution of cluster galaxies and their associated sub-halos is not well reproduced in all CDM simulations including the Illustris-TNG as demonstrated in \citet{Natarajan+2017}. Therefore, if this additional mismatch is attended to in simulations, it would result in the production of more massive sub-halos closer to the cluster center, which would in turn also increase the GGSL. More accurately reproducing observed cluster galaxy properties within CDM is an on-going challenge and project for all simulators and this involves implementing more physical models for star formation, black hole growth and feedback \citep{Ragagnin+2022}.

We conclude that sub-halos possessing internal density profiles given by Eqn.~\ref{eqn:gnfw_density_profile} and inner slopes $\gamma \sim 3$ can resolve the GGSL discrepancy. This enhancement of the GGSL probability occurs due to the large number of sub-halos that become super-critical for lensing in cluster environments. The increased GGSL probability results from the steepened profile slope and the additional boost provided by the larger scale mass components in the cluster (discussed further in the Methods section). Note that NFW halos in field environments with $\gamma \sim 1$ are sub-critical, and cannot themselves generate multiple images. The super-critical densities on these scales in CDM come from the added contribution of baryonic matter. If one invokes a baryonic solution that increases the central density of a halo by rearranging stellar mass instead of dark matter, our results show that the required outcome must reproduce a mass density profile given by Eqn.~\ref{eqn:gnfw_density_profile} with $\gamma > 2.5$. However, no known mechanisms cause this level of steepening on the scales required to explain the GGSL discrepancy\citep{Heinze+2023}. 

The steepening of inner dark matter profiles via multiple physical processes have been explored previously including the following: (i) accretion of dark matter onto central black holes which achieves $\gamma \simeq 2.3$ \citep{Gondolo+2000} on scales $< 1\,{\rm pc}$; (ii) formation of dark matter spikes around Intermediate Mass Black Holes results in $\gamma \simeq 2.25$ and (iii) the adiabatic assembly of black holes resulting in dark matter mounds \citep{Bertone+2024}. None of the above steepen NFW profiles on scales of interest for GGSL. Examining the impact of baryons concentrated in the inner regions of dark matter halos on the inner density profiles, \citet{DiCintio+2014a,DiCintio+2014b} derived a relation between the inner density slope and the stellar-to-halo mass ratio for simulated galaxies in the MUGS and MaGICC projects, finding a steepening of $\gamma < 1.5$. This is insufficient to bridge the GGSL discrepancy.\citet{Schaller+2015,Schaller+2015a} report steepening of the NFW to $\gamma < 2.0$ in the EAGLE suite of cosmological simulations, while \citet{Heinze+2023} found a steepening to $\gamma = 2$ when fitting subhalos in the TNG50 simulation, once again, these slopes are insufficient to produce the enhancement needed in the GGSL probability. \citet{Moline+2017,Moline+2022} have shown that the central concentration of cluster subhalos is enhanced by virtue of them residing in denser environments by a factor of 2-3 compared to the standard concentration used in the NFW profile in $\Lambda$CDM model. We find that increasing the concentration by this factor of 2-3 still does not help alleviate the GGSL discrepancy. For details see discussion and plot in Section \ref{sec:impact_concentration_ggsl} of the Appendix.

\begin{figure}[ht]
    \centering
    \includegraphics[width=1.0\textwidth]{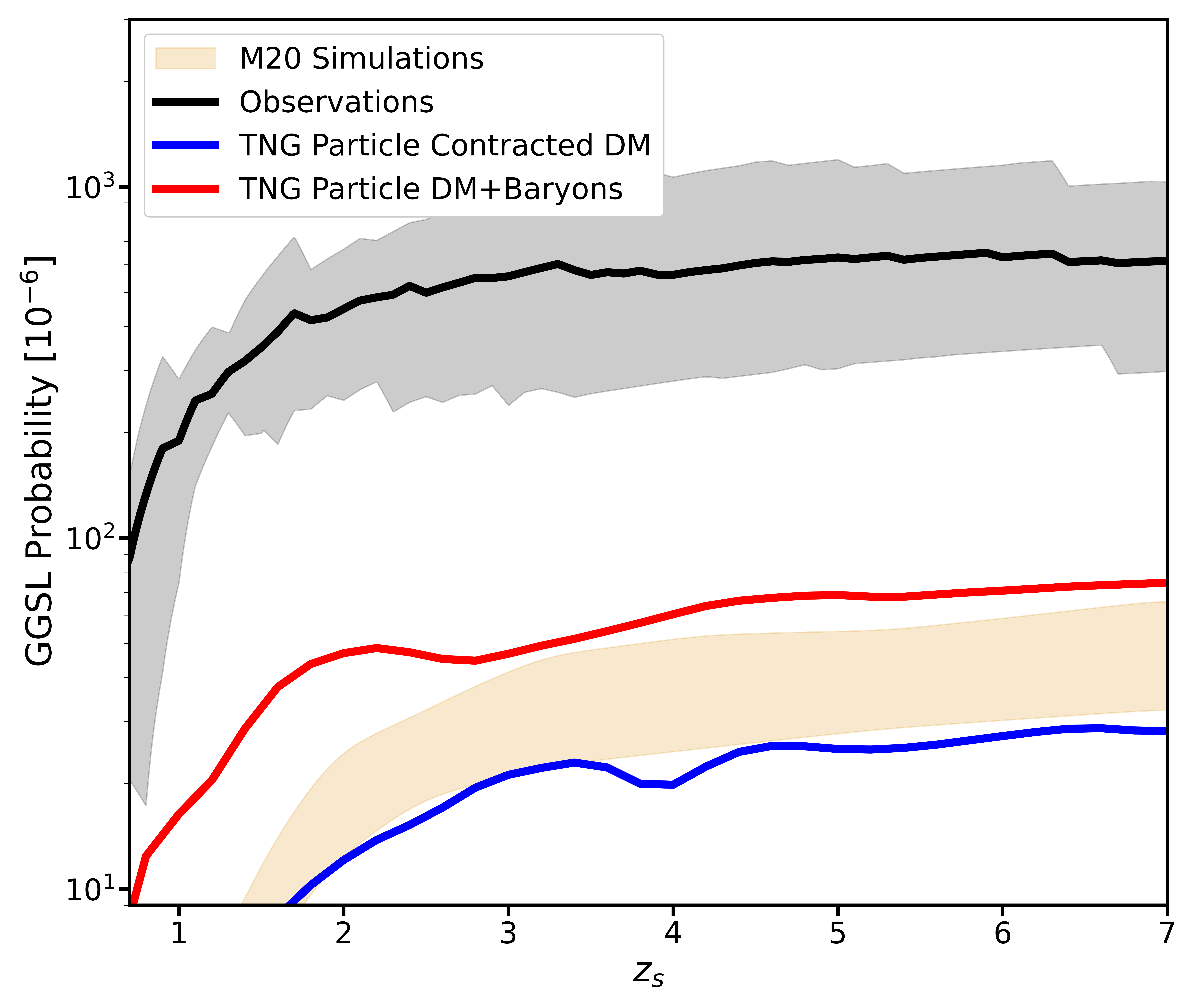}
    \caption{The GGSL probability as a function of source redshift computed by smoothing the particle data from the TNG-300 simulations for observed clusters and their mass-matched simulated analogs. The blue curve is the GGSL computed from the analogs taken from the TNG300 with dark matter particles that includes the effects of the induced adiabatic contraction due to the baryons and in red, we show the GGSL calculated explicitly including all the components - the dark matter, gas, stars and black holes in the TNG-300 simulated cluster analogs. 
     }
    \label{fig:ggsl_tng_particle}
\end{figure}

\begin{figure}[ht]
    \centering
    \includegraphics[width=1.0\textwidth]{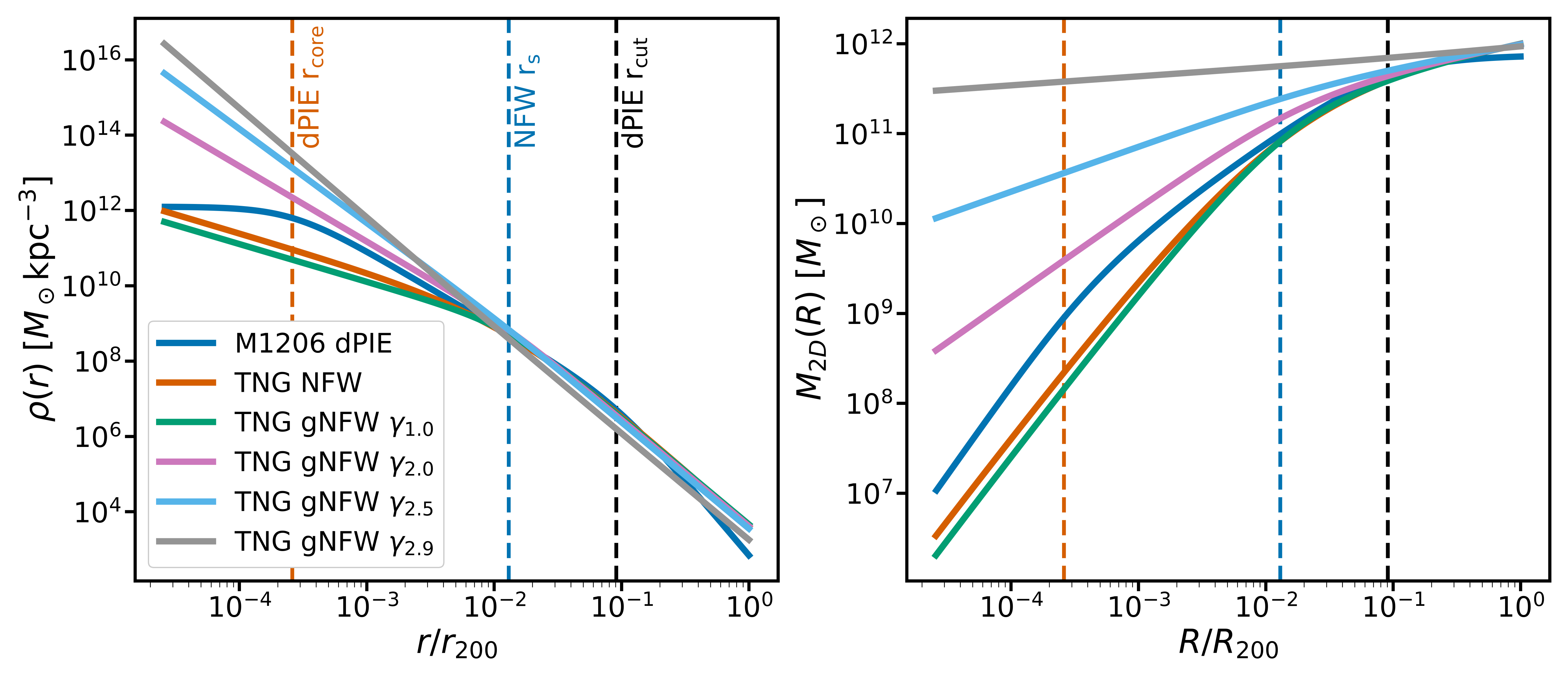}
    \caption{Changes in inner density slope:  we show the changes in the density profile shape and the corresponding change in the enclosed projected mass (2D mass) of a fiducial cluster subhalo in the observed lensing cluster M1206 (that is part of our sample) and a subhalo with matching $M_{200}$ from the simulated cluster analog. Note that steepening the inner logarithmic slope $\gamma$ from $1$ to $2.9$ results in an enhancement of the central density of the sub-halo by four orders of magnitude within 0.1 $r_{200}$.}
    \label{fig:m1206_rho_m2d}
\end{figure}

\begin{figure}[ht]
    \centering
    \includegraphics[width=1.0\textwidth]{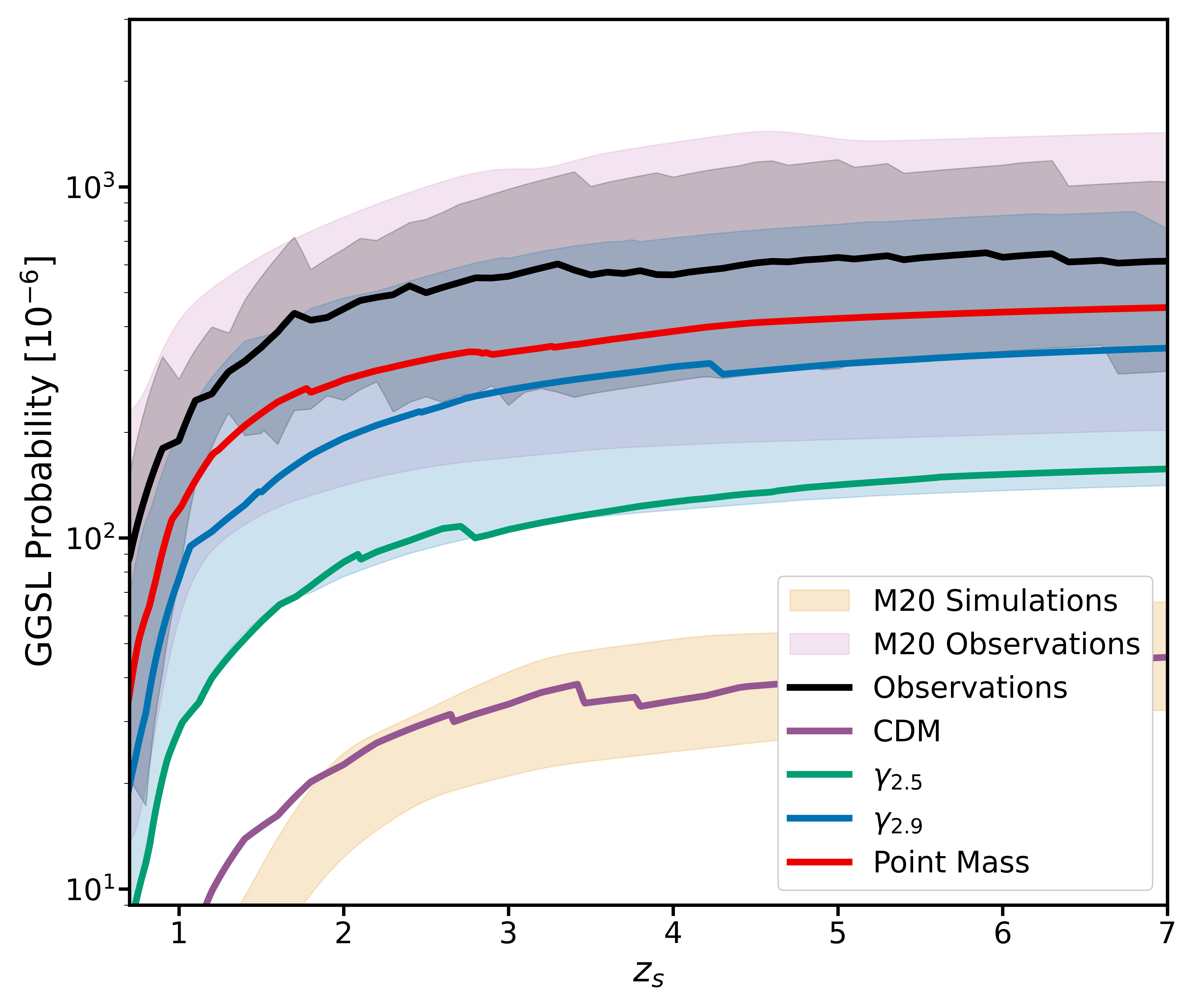}
    \caption{The GGSL probability as a function of source redshift for observed clusters, simulated mass-matched analogs in CDM and c-SIDM. The mean GGSL probability of our sample is shown in black and its range are depicted with a gray band. The median GGSL probability for the CDM TNG-300 simulation suite is shown by a purple line; the blue and green lines show the median GGSL probability of the mimicked CDM subhalos undergoing core collapse; we show the 95\% percentile with enhanced central density by the blue band; and the GGSL for the final collapsed subhalos in the limit as point masses is plotted with the red line. Pink and orange bands signify the range of GGSL probability for observed clusters and simulations, respectively from M20.}
    \label{fig:ggsl_sidm}
\end{figure}

\begin{figure}[ht]
    \centering
    \includegraphics[width=1.0\textwidth]{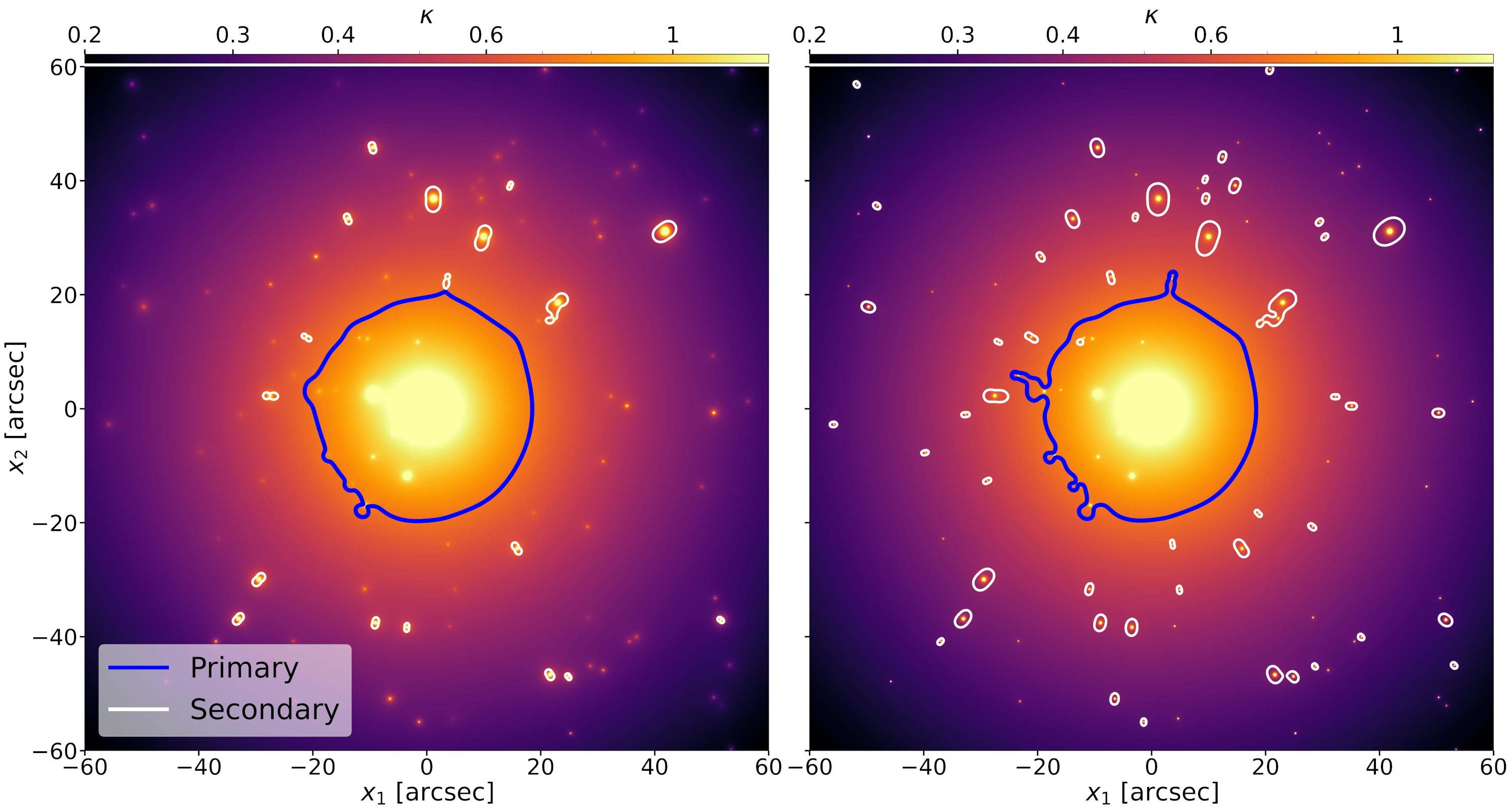}
    \caption{CDM and c-SIDM critical lines of a $10^{15} M_\odot$ halo found in TNG300 simulation. The left figure shows the critical lines of the halo modeled with CDM profiles, while the right shows the steeper $\gamma$ profiles. The critical lines of these steepened subhalos increase in area and number as more subhalos become supercritical, while the subhalo critical lines close enough to the center merge with the primary critical line, creating visible distortions on its contours.}
    \label{fig:tng_critical_lines}
\end{figure}


\section{Conclusions \& Discussion}

Our simulations quantify the central density profiles of dark matter halos required to reproduce the observed GGSL probability of observed cluster lenses. Our results show that the galaxies in dense cluster environments have higher central densities than those produced in large-scale cosmological simulations of these structures. If the discrepancy arises entirely from systematic issues present in the simulations, it must enter at a level that increases the central densities of halos by four orders of magnitude as seen in Fig.~\ref{fig:m1206_rho_m2d}. \citet{Moline+2017} have shown that the central concentration of simulated cluster subhalos is enhanced by virtue of their being in denser environments by a factor of 2-3 compared to the standard concentration used in the NFW profile in $\Lambda$CDM model. We find that increasing the concentration by this factor of 2-3 still does not help alleviate the GGSL discrepancy. And in more recent work, using even higher resolution $\Lambda$CDM simulations, \citep{Moline+2022} find that even larger enhancements that depend on the distance of the subhalo from the cluster center result, which is still insufficient to account fully account for the GGSL discrepancy. However, we caution that these analysis are derived from dark-matter only simulations that do not include baryonic feedback processes that re-distribute matter on the smallest scales. 

Therefore, the discrepancy between measurements and simulations based on CDM illustrated by Fig.~\ref{fig:ggsl_sidm} imply that current state-of-the-art cosmological simulations do not accurately predict the internal structure of dark matter halos on scales $10^{10} - 10^{12} M_{\odot}\,{h^{-1}}$. This scenario implies that these simulations have missing physics associated with the galaxy formation and feedback mechanisms in clusters, or they could point to different properties of the dark matter than predicted by CDM. 

The calculation of the GGSL itself does not depend on dark matter physics. However, one can interpret the implied properties of halos within the context of dark matter physics, and in particular SIDM, which predicts an enhanced central densities for subhalos through core collapse. To date, this process has mainly been studied in less massive halos outside of galaxy clusters. 

The process of core collapse depends on both the internal structure of a dark matter halo and the amplitude of the self-interaction cross section of a halo \citep[e.g.][]{Yang+2022}. As shown by \citet{Kaplinghat+2016}, halos with different masses probe the SIDM cross section at different relative velocities. For the cluster subhalos considered in this work, the relevant velocity scales are $\sim 50 -100 \ \rm{km} \rm{s^{-1}}$, and thus the GGSL within the context of SIDM requires a self-interaction cross section on these velocity scales high enough to drive core collapse. 

A natural question in this context is whether a viable SIDM cross-section exists that can drive core collapse on these scales without violating constraints from other observational probes. At velocities $\sim 1,000 \ \rm{km} \ \rm{s^{-1}}$ stringent upper limits on the cross-section amplitude $\sigma \lesssim 1 \rm{cm^2} \rm{g^{-1}}$ \citep{Peter+2013,Sagunski+2021} preclude values high enough to cause core collapse in cluster sub-halos within the age of the Universe. However, inelastic scattering cross-sections can accelerate the onset of core collapse, and possibly result in core collapse within the age of the Universe on these scales with an interaction cross section $\mathcal{O}\left(1 \rm{cm^2} \rm{g^{-1}}\right)$ \citep{Essig+2019,Huo+2020,ONeil+2023}. 

Alternatively, velocity-dependent cross-sections can increase the cross-section amplitude on the velocity scales of interest, especially if a mechanism specific to cluster environments accelerates the onset of core collapse relative to galactic environments. Velocity-dependent cross sections arise naturally in SIDM models in which dark matter particle with mass $m_{\chi}$ interacts through a light force carrier $m_{\phi}$ \citep{Vogelsberger+2016,Tulin+2013,Colquhoun+2021,Yang+2022}. At low speeds ($v \propto \frac{m_{\phi}}{m_{\chi}}$), these cross sections typically scale inversely with relative velocity to a power $\alpha$ that varies from close to zero for certain models with a weak potential (i.e. those in the Born regime), to values of $\alpha$ the range $-1$ to $-4$ in the semi-classical or classical scattering regimes \citep{Tulin+2013,Colquhoun+2021}. In such models, an immediate consequence of an appreciable cross section on the scales of cluster sub-halos is that the cross-section amplitude on lower mass scales is enhanced by a factor $\left(\frac{v}{100\rm{\ km \ \rm{s^{-1}}}}\right)^{-\alpha}$, corresponding to a factor 100-1000 increase on the scales of dwarf galaxies $v \sim 10-50 \ \rm{km} \rm{s^{-1}}$, and by an even larger factor on the scales probed by galaxy-galaxy strong lensing \citep{Minor+2021,Gilman+2021,Gilman+2023}. Alternatively, a cross-section with a pronounced resonance precisely at the velocity scales relevant for cluster sub-halos \citep[e.g.][]{Tulin+2013b,Chu+2019,Gilman+2023} could drive collapse in these structures without violating constraints from other scales, although one could argue that these solutions are disfavored because they require fine tuning of the particle physics model. 

To understand whether SIDM presents a viable resolution to the GGSL tension requires high-resolution hydrodynamical simulations of SIDM on galaxy cluster scales with sufficient resolution to resolve the central density profiles of cluster sub-halos as they collapse. \citet{Ragagnin+2024} recently simulated SIDM on cluster scales and reported a central density enhancement of $\sim 20\%$ in cluster subhalos, but their simulations did not have sufficient resolution to study core collapse, and thus this phenomenon on cluster scales remains unexplored. However, if one assumes that the GGSL tension reported by M20 stems entirely from the internal properties of cluster sub-halos, SIDM models offer a viable solution through core collapse. Whether an SIDM solution on the scales of cluster sub-halos can evade stringent constraints from other astrophysical probes remains an open question.


\section*{Acknowledgments}
The authors acknowledge useful discussions with Massimo Meneghetti, Guillaume Mahler, Mathilde Jauzac, Yarone Tokayer, Simon Birrer, Urmila Chadayammuri, and Elena Rasia. The authors are grateful to the anonymous referee for their constructive feedback. PN gratefully acknowledges funding from the Department of Energy grant DE-SC0017660. DG acknowledges support for this work provided by The Brinson Foundation through a Brinson Prize Fellowship grant.

\vspace{5mm}

\software{astropy \citep{TheAstropyCollaboration+2022}, numpy \citep{Harris+2020}
scikit-image \citep{VanDerWalt+2014}, lenstronomy \citep{Birrer+2018,Birrer+2021}
}

\appendix

\section{Impact of projection effects on GGSL}
\begin{figure}[ht]
    \centering
    \includegraphics[width=1.0\textwidth]{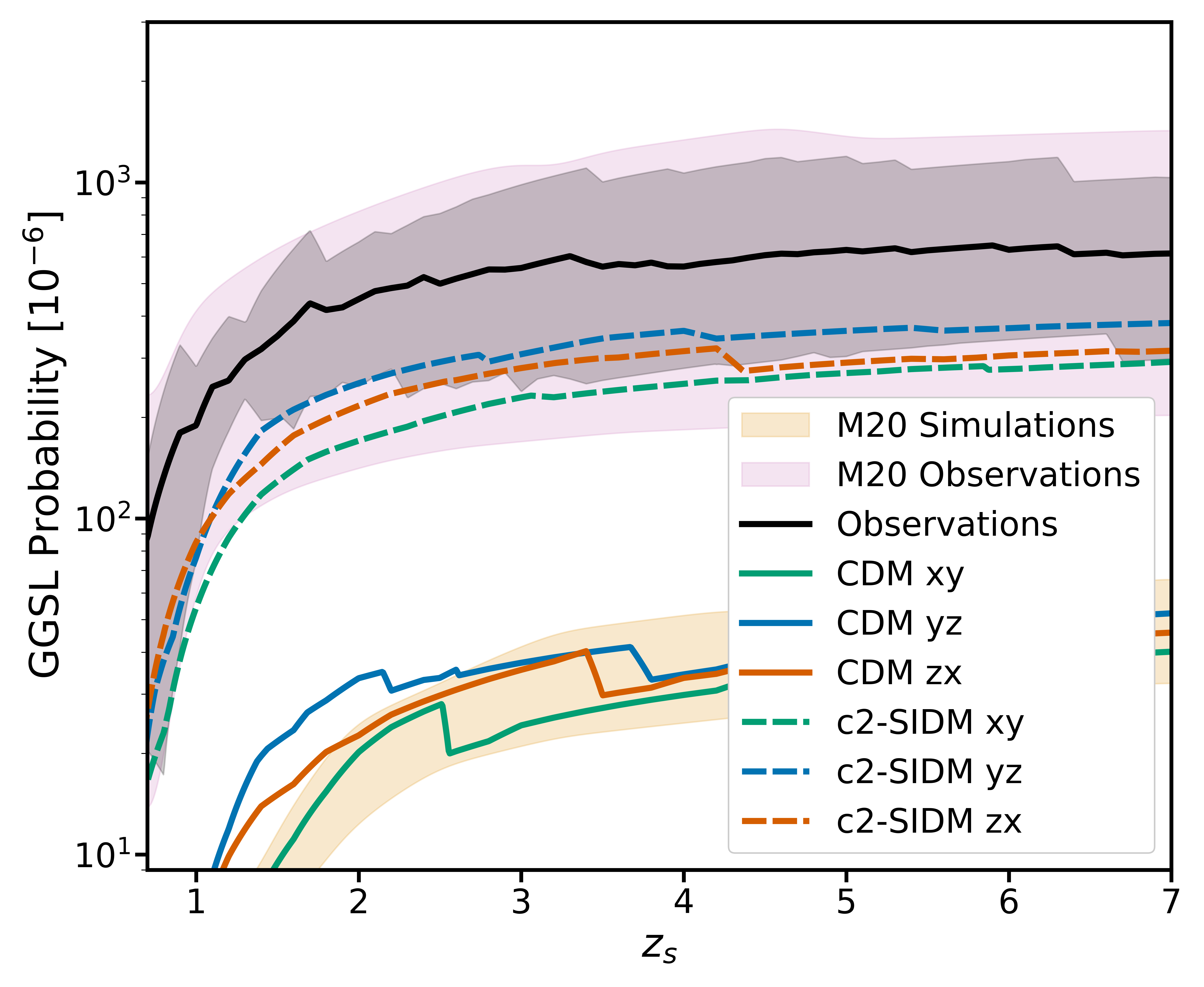}
    \caption{Impact of projection effects: change in GGSL probability as a function of source redshift along the three independent projections. Once again, the mean GGSL probability of our sample is shown in black and its range as a gray band. The GGSL probability for each independent projection for the selected CDM TNG300 analog is shown by solid lines in blue, green and red; the dashed lines with corresponding colors show the GGSL probability of the steepened c2-SIDM.}
    \label{fig:ggsl_projection}
\end{figure}

We investigate the impact of projection effects on GGSL from the simulated analogs and find it to be negligible. For a simulated analog in our sample, we compute the GGSL probability for each independent projection for both CDM and the steepened version (labeled as c2-SIDM), and show the results in Fig.~\ref{fig:ggsl_projection}. The highest computed GGSL probability at redshift $z = 7$ differs from the lowest by less than 20\% for CDM and by $\sim$ 26\% for c2-SIDM, insufficient in and of itself to account on its own for the reported GGSL discrepancy. Therefore, projection effects cannot alleviate the GGSL discrepancy due exclusively one of the projections, for some clusters align in the direction of elongation of the cluster.

\begin{figure}[ht]
    \centering
    \includegraphics[width=1.0\textwidth]{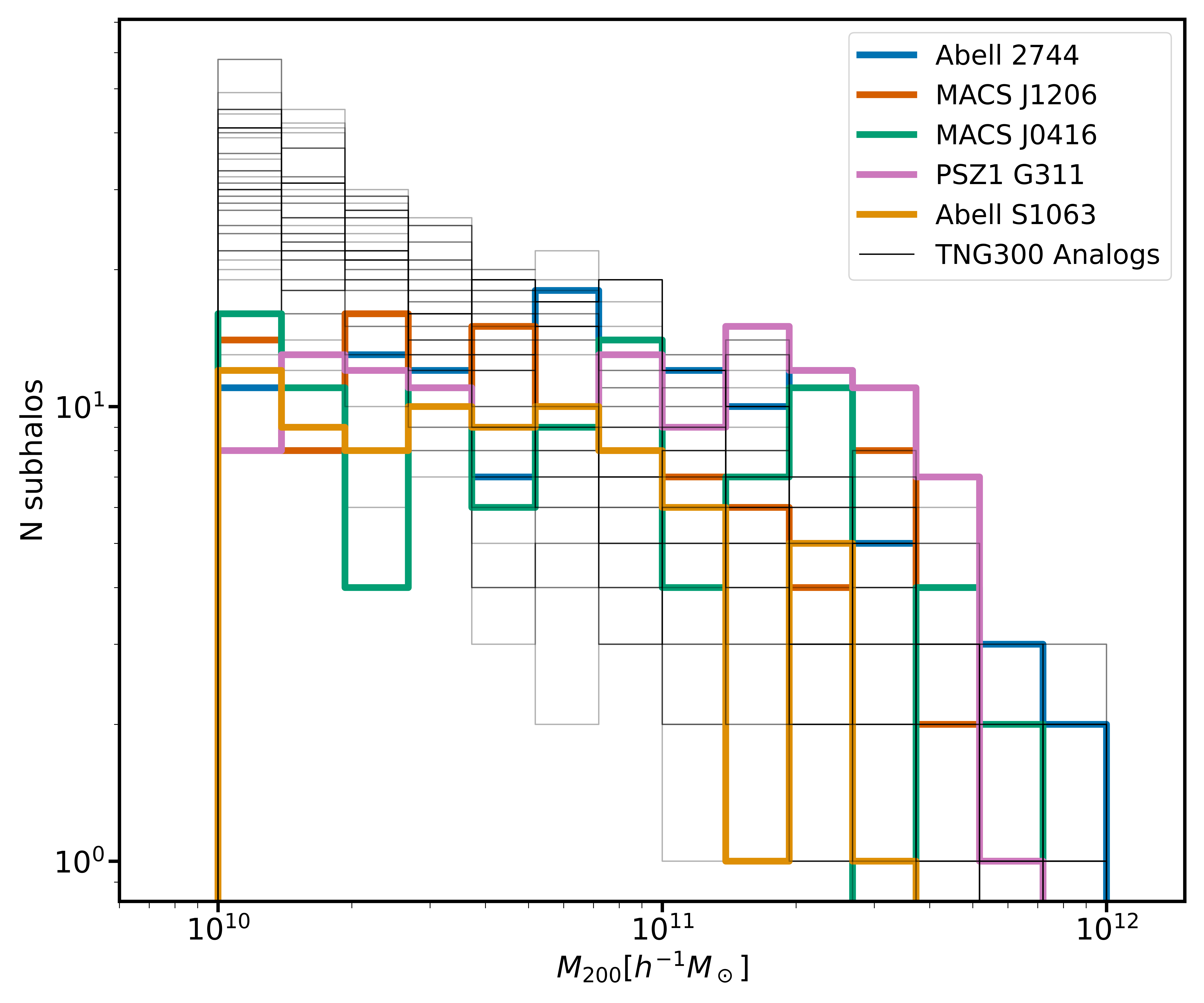}
    \caption{The subhalo mass function of observed clusters and their simulated CDM analogs in the TNG300 box: The solid colored lines are forthe observed cluster sample, while gray lines show the subhalo mass distributions of the simulated CDM analogs.}
    \label{fig:shmf_all}
\end{figure}

\section{Impact of ellipticity on GGSL}\label{sec:ellipticity}

\begin{figure}[ht]
    \centering
    \includegraphics[width=0.49\textwidth]{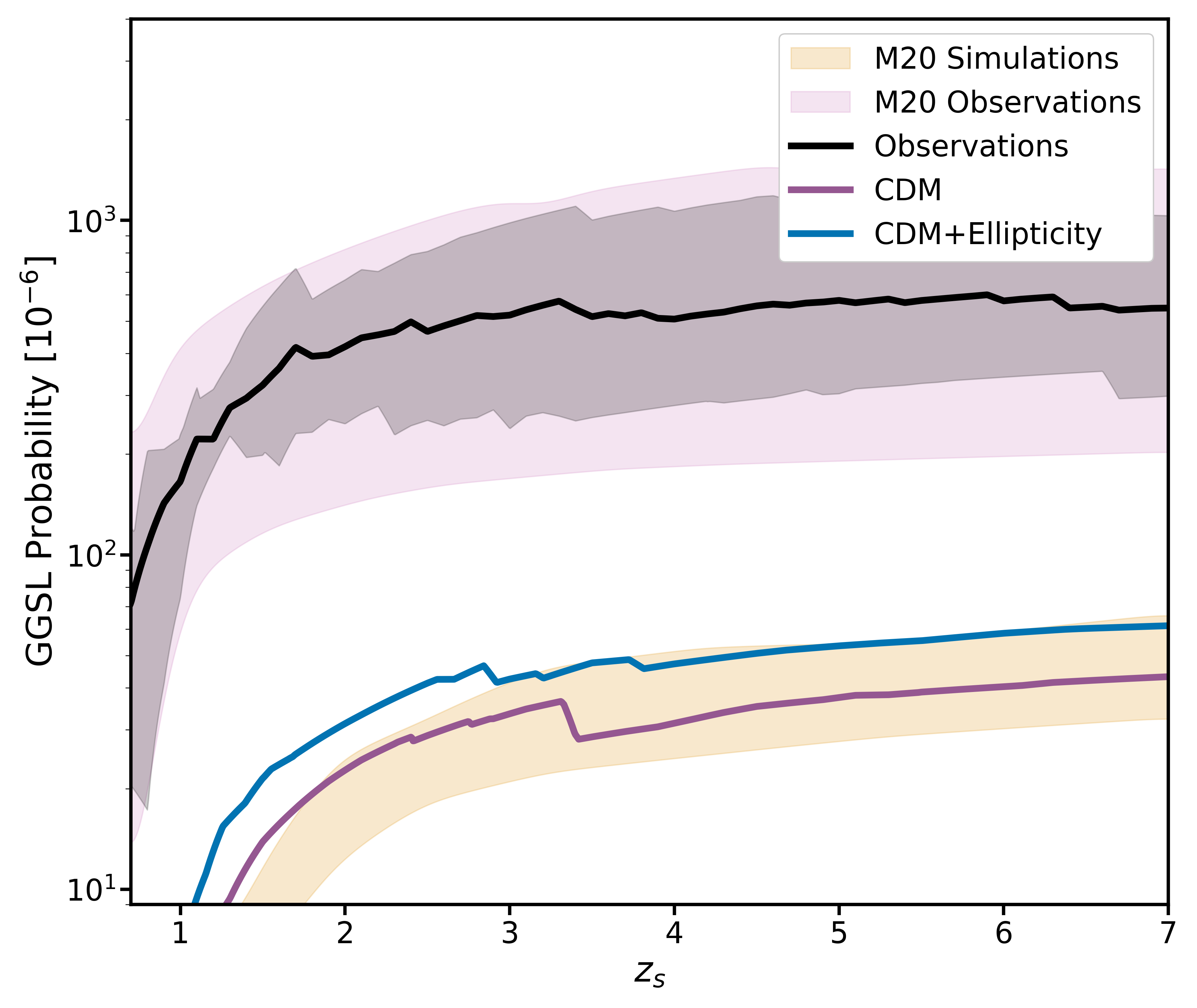}
    \includegraphics[width=0.49\textwidth]{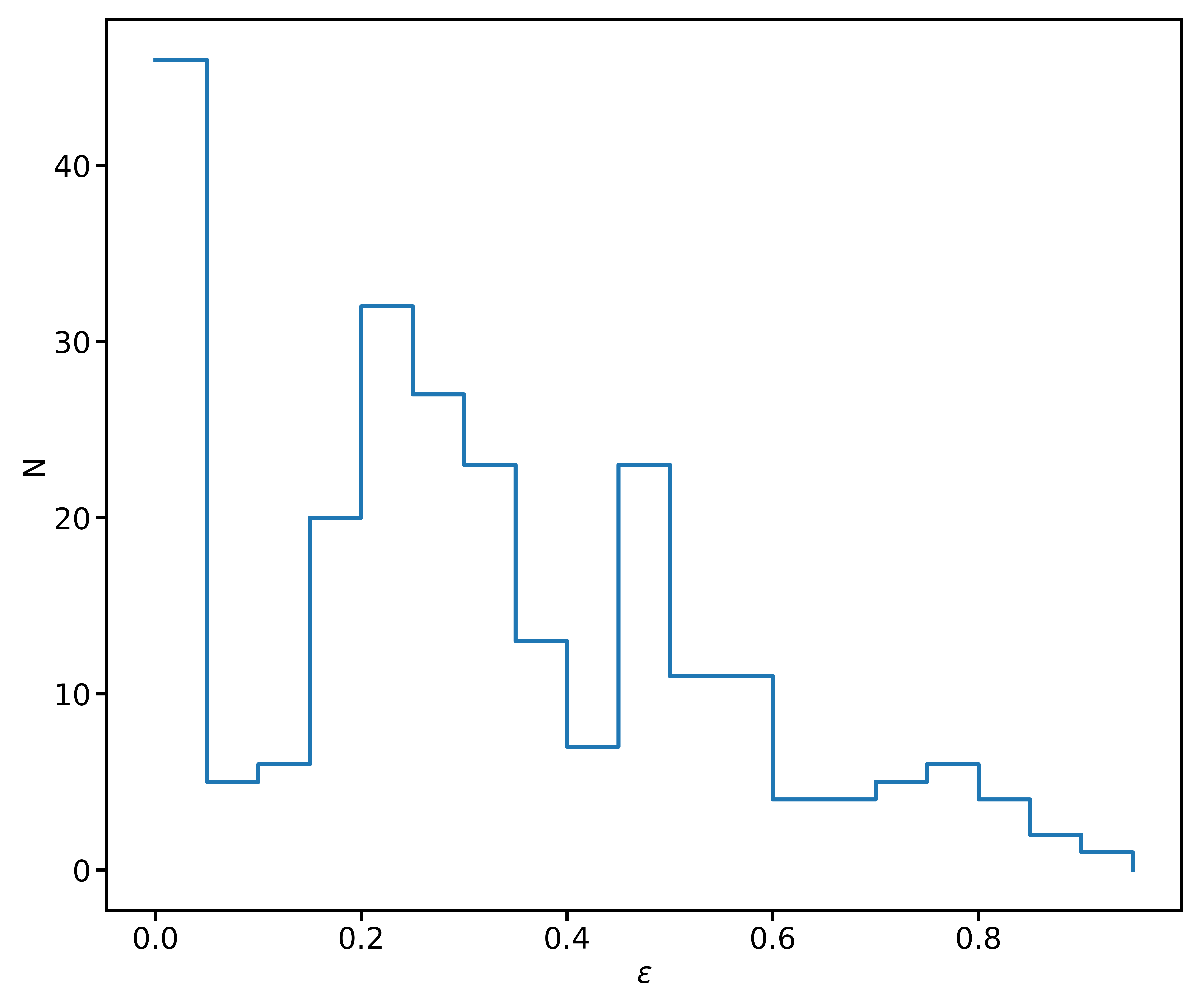}
    \caption{Left: Comparison of the GGSL probability for spherical and elliptical subhalos: we compute the GGSL probability for the CDM TNG300 simulated analogs as spherical NFW profiles; and overplot the calculation done with elliptical NFW profiles wherein the ellipticity is assigned to subhalos by drawing uniformly from Abell 2744 ellipticity distribution.}
    \label{fig:ggsl_ellipticity}
\end{figure}

We investigate the effect of subhalo ellipticity on our GGSL calculations and also find that it is insufficient to fully account for the GGSL discrepancy. When computing the GGSL probability from simulations are described above, we first initialize a spherical NFW profile in \texttt{lenstronomy} for each subhalo in the mass matched simulated cluster analog. We note that  there are no observational constraints on the ellipticity of dark matter subhalos. Therefore, all the \texttt{Lenstool} parametric model files our observed sample typically adopt spherically averaged profiles. However, if we assume that the ellipticity of the light distribution in an observed cluster member galaxy is a proxy for the ellipticity of its associated dark matter halo, as was done for the case Abell 2744, we can then construct an ellipticity distribution for the subhalo population. The \texttt{Lenstool} model for Abell 2744 adopted this ellipticity distribution as a prior, and it is shown in the left panel of Fig.~\ref{fig:ggsl_ellipticity}. 

The ellipticity $\hat{\epsilon}$ of a projected mass distribution with semi major axis A and semi minor axis B are defined by
\begin{equation}
    \hat{\epsilon} = \frac{A^2 - B^2}{A^2 + B^2}
\end{equation}

Further details can be found in \cite[Appendix A.5]{Eliasdottir+2007} and references therein.

We adopt this observationally derived ellipticity distribution for subhalos in Abell 2744 and draw uniformly from it to initialize elliptical NFW profiles in \texttt{lenstronomy} for each simulated analog and then compute the GGSL probability for the rest of our observed sample. The result can be seen in the left panel of Fig.~\ref{fig:ggsl_ellipticity}. We notice a slight increase in the GGSL probability by a factor of about 1.4 when the effects of subhalo ellipticity are included. However, the effect of ellipticity appears to be insignificant in and of itself to account for the observed GGSL discrepancy between simulations and observed cluster-lenses. 

\section{Computing GGSL from particle data for cluster analogs in Illustris-TNG}\label{sec:ggsl_particle}

We validate our NFW profile fits and the computation of the GGSL probability for mass-matched simulated analogs from the Illustris TNG300 simulation by computing the GGSL probability directly from the TNG300 particle data. To achieve this, we derive the deflection angle maps by tracing rays through the lens based on the projected mass distributions of these simulated halos, similar to the procedure employed in \citet{Meneghetti+2020} and implemented in the publicly available code \texttt{Py-SPHViewer}\footnote{Publicly available at \url{https://github.com/alejandrobll/py-sphviewer}}. We slice a box around the cluster with a depth of 20 Mpc and use a Smoothed-Particle-Hydrodynamics scheme to create a convergence map, projecting each particle species separately (dark matter, gas, stars, and black holes) to achieve the final convergence map.

We compute the GGSL probability for the same mass-matched simulated analogs from TNG300 particle data for two distinct cases:
\begin{itemize}
    \item From the TNG300-1 simulation that includes dark matter and baryons particles, we compute the GGSL probability by creating a smoothed convergence map from including only the dark matter particles in an effort to capture the effect of baryonic-induced adiabatic contraction on the dark matter halos.
    \item From the TNG300-1 simulation that includes dark matter and baryons particles, we compute the GGSL probability by creating a smoothed convergence map from including dark matter and baryons (gas, stars and black holes) particles.
\end{itemize}

\begin{figure}[ht]
    \centering
    \includegraphics[width=1.0\textwidth]{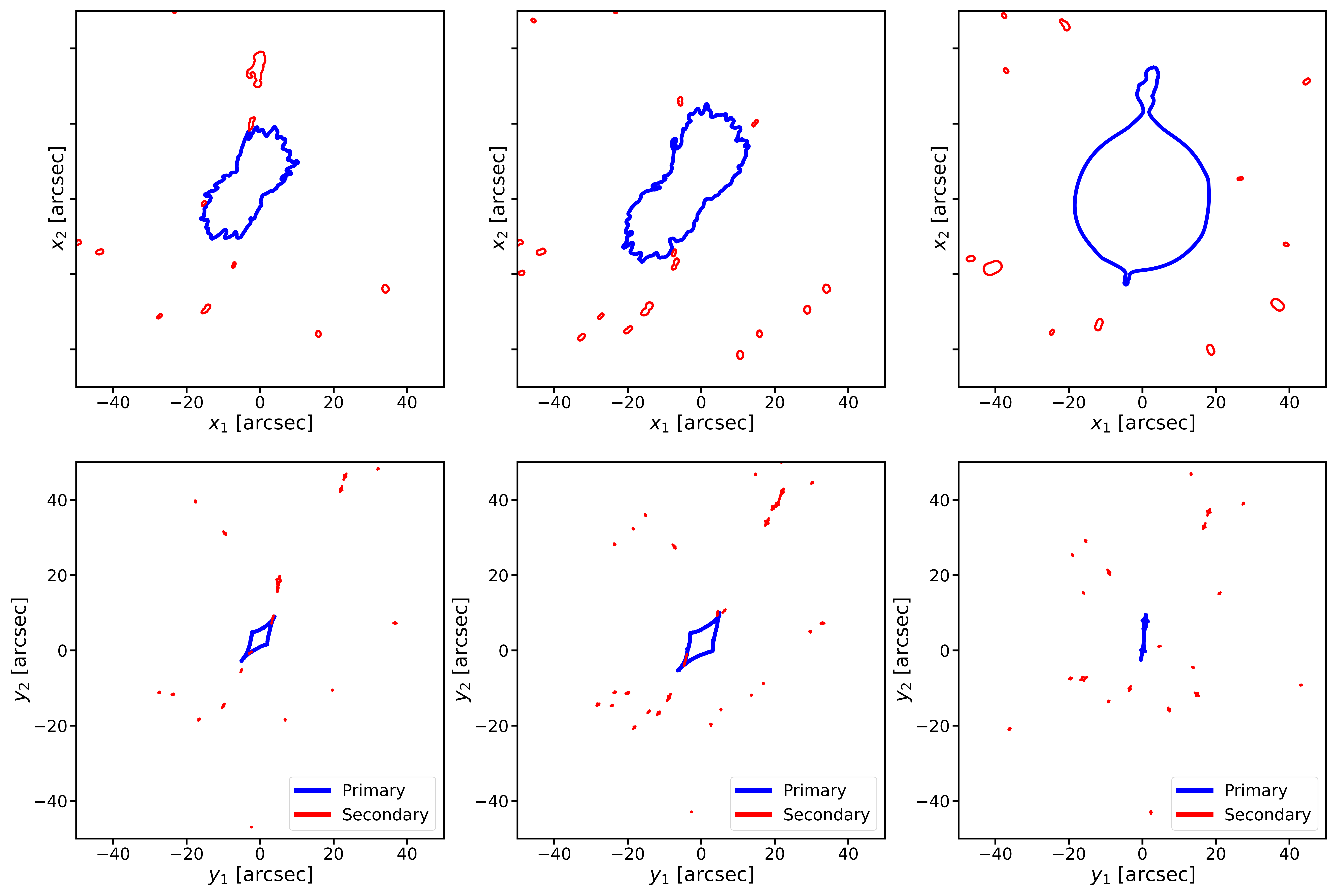}
    \caption{The lensing critical lines, shown in the top panels, and projected caustics, shown in the bottom panels, for the same TNG300 simulated analog. In the left panel and central panel, of the figure we plot the critical lines were computed by smoothing and projecting the particle data as described above, only including dark matter particles (left panel) and including baryonic particles (central panel). The right panel shows the critical lines computed from our NFW fits.}
    \label{fig:tng_particle_caustics}
\end{figure}

The resulting lensing critical lines and caustics of an observed cluster from our sample and a mass-matched simulated analog can be seen in Fig. \ref{fig:tng_particle_caustics}.
We note that the Einstein radius, computed as specified in Section \ref{sec:calculation_ggsl}, of the primary critical line computed by fitting NFW profiles is 19 arcseconds, while for the one computed from the particle data is 18 arcseconds. The results of GGSL probability computation are shown in Fig. \ref{fig:ggsl_tng_particle}. We note that while including the effects of baryon-driven contraction; and all the baryonic components explicitly - namely, stars, gas, dust and black holes results in an enhancement of the GGSL probability, however these are still insufficient to bridge the discrepancy with observations.

\section{Impact of subhalo concentration on GGSL}\label{sec:impact_concentration_ggsl}

While \citet{Tokayer+2024} found it insufficient to bridge the GGSL discrepancy, we investigate the impact of the NFW concentration parameter on the subhalo properties of interest, in this case the GGSL probability. 

\begin{figure}[ht]
    \centering
    \includegraphics[width=1.0\textwidth]{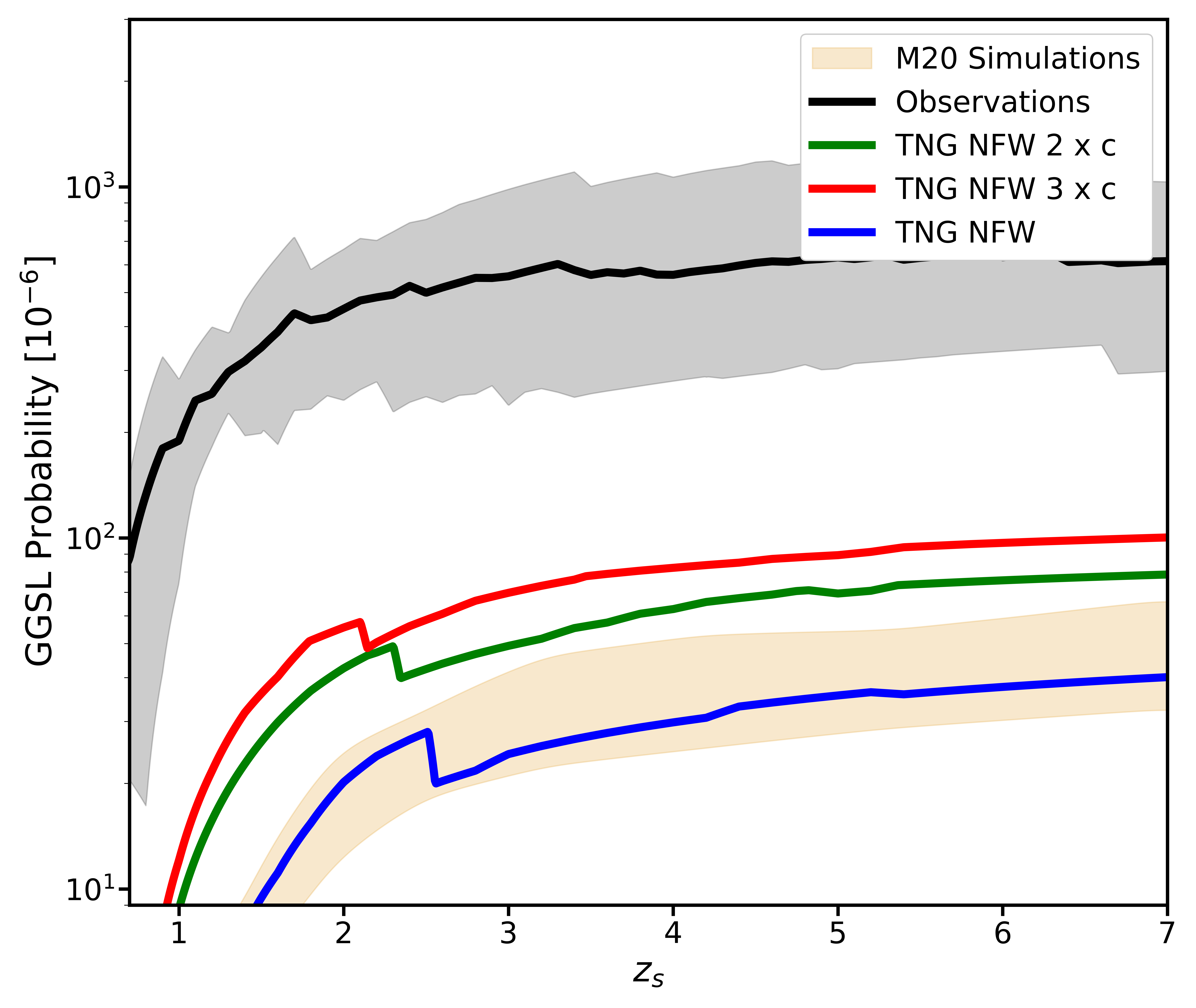}
    \caption{The GGSL probability computed for our simulated mass-matched Illustris subhalos including an enhancement in the NFW concentration parameter by factors of 2-3, as initially reported in \citep{Moline+2017}. This increase does not alleviate the GGSL discrepancy.}
    \label{fig:ggsl_tng_concentration_multiples}
\end{figure}

As shown by \citet{Moline+2022} from the results of analyzing their high-resolution dark-matter only Uchuu simulation, subhalos may be significantly more concentrated when residing inside massive hosts, specially those located in the inner regions have a significantly higher concentration than those in the outer regions. Using the Uchuu suite of dark-matter only N-body cosmological simulations, \citet{Moline+2022} quantified subhalo concentrations using the parameter $c_\mathrm{v}$, defined as the mean density found at the radius of maximum circular velocity $R_\mathrm{Vmax}$, and found it to be in the range of $10^4$ to $10^5$ (for the subhalos considered here), decreasing for more massive subhalos. We note that the more familiar NFW concentration can be directly related to $c_\mathrm{v}$ \citep{Moline+2017} including a dependence on the projected distance from the cluster center, via: 

\begin{equation}
c_\mathrm{v} = \left( \frac{c_\mathrm{NFW}}{2.163} \right)^3 \frac{f(R_\mathrm{max}/r_\mathrm{s})}{f(c_\mathrm{NFW})} \Delta
\end{equation}
where $f(x) = \ln{(1+x)} - x/(1+x)$.
The concentrations that we obtain for simulated subhalos are also higher than for comparable mass isolated halos, and these tend to be outliers in the concentration-Mass (c-M) relation in $\Lambda$CDM. However, as we have seen this still does not translate into a high enough value for the GGSL probability to account for discrepancy as shown above with halo catalogs and also with the particle data. As shown by \citet{Moline+2017} from the results of analyzing their high-resolution dark-matter only Uchuu simulation, subhalos may be significantly more concentrated when residing inside massive hosts, specially those located in the inner regions have a significantly higher concentration than those in the outer regions, enhanced by up to a factor of $\sim 3$ for less massive subhalos and $\sim 1.5$ for most massive subhalos. We note that these enhancement factors are still insufficient to account for the GGSL probability in observed lensing clusters as shown in Fig. \ref{fig:ggsl_tng_concentration_multiples}.

As shown in Fig. \ref{fig:m1206_rho_m2d} above, we need concentration of mass well inside the NFW scale radius ($r_s$) on scales less than $0.01 \times R_200$ to impact the GGSL probability and increase it significantly. We calculate $c_\mathrm{v}$, and subsequently $c_\mathrm{NFW}$,for all subhalos in our simulated analogs by following the parametrization given in \citet{Moline+2022}:

\begin{equation}
    c_\mathrm{v}(V_\mathrm{max}, x_\mathrm{sub}) = c_0 \left[ 1 + \sum_{i=1}^3 \left[ a_i \log_{10}\left( \frac{V_\mathrm{max}}{\mathrm{km}\, \mathrm{s}^{-1}} \right) \right]^i \right] \times \left[ 1 + b \log_{10}(x_\mathrm{sub}) \right]
\end{equation}
where $x_\mathrm{sub} = r_\mathrm{sub}/R_\mathrm{vir,h}$ is the ratio of subhalo distance to the host halo center, $c_0 = 1.12 \times 10^5$, $a_i = {-0.9512, -0.5538, -0.3221}$, and $b = -1.7828$. 

\begin{figure}[ht]
    \centering
    \includegraphics[width=0.37\textwidth]{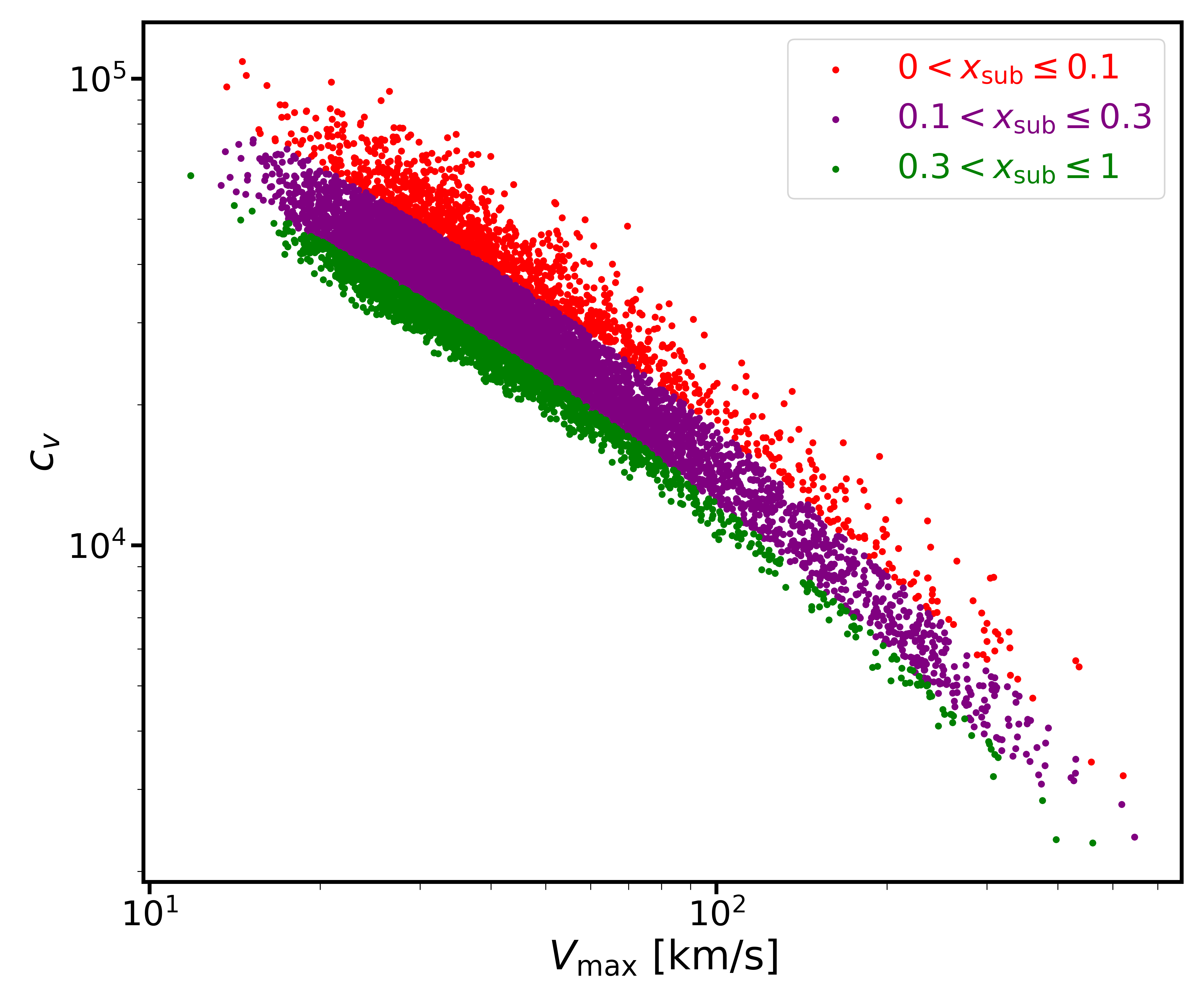}
    \includegraphics[width=0.61\textwidth]{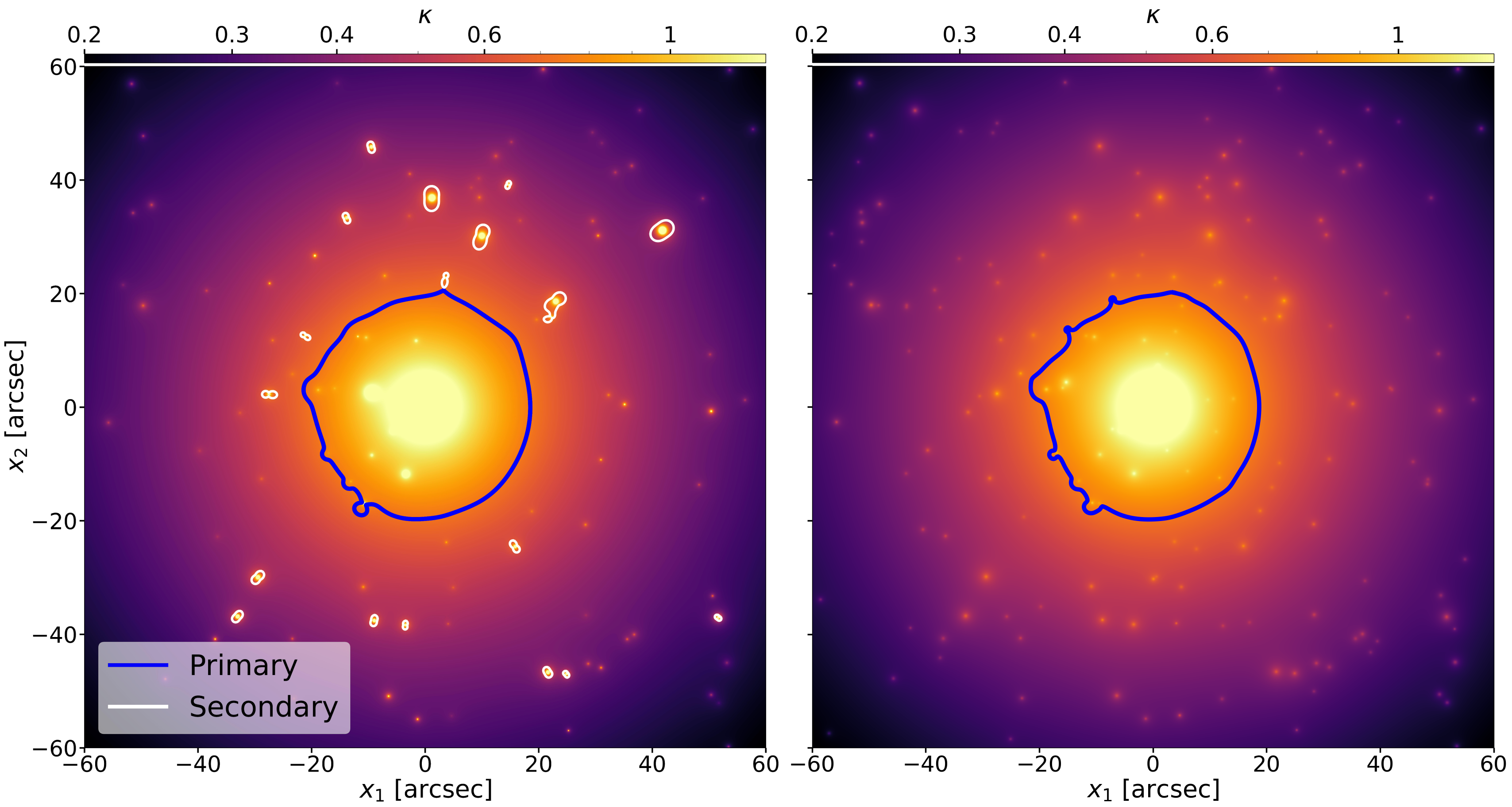}
    \caption{Left: The values of $c_v$'s computed from the $c_\mathrm{NFW}$'s for our Illustris sub-halos as a function of $v_{\rm max}$ using the conversion equation that includes the dependence on projected distance from the cluster center explicitly. We note that these distributions match those obtained by \citet{Moline+2022} in Fig. 6 of their paper. Therefore, the claimed enhancement in concentration from their higher resolution simulation Uchuu is also seen in Illustris TNG-300 subhalos. Middle and right: The convergence map of a simulated analog in TNG contrasting the critical lines obtained fitting subhaloes with a concentration parameter used in this work (middle) to the ones obtained using \citet{Moline+2022} concentration parametrization (right).}
    \label{fig:moline_cv}
\end{figure}

In the left panel of Fig. \ref{fig:moline_cv}, we show the values of $c_v$ obtained for the mass-matched simulated analog Illustris subhalos given their $c_\mathrm{NFW}$. It is clearly seen to be very similar to the distribution derived by \citet{Moline+2022} in Figure 6 of their paper. In the right and middle panels, we show the critical lines of a TNG simulated analog obtained from fitting the subhalos concentration parameter using the parametrization provided in \citet{Moline+2022}, and contrast it to the critical lines obtained from the concentration adopted in this work, respectively. Therefore, our Illustris subhalos are also as over-concentrated as the subhalos in their Uchuu simulation. However, as we have already shown in this work, the Illustris mass matched analog subhalos fail to account for the GGSL discrepancy while using both the halo catalogs as well as the particle data directly. Redefining the concentration parameter does not alter their contribution to GGSL.

Therefore, we demonstrate that neither projection effects, ellipticity, nor the concentration can fully alleviate the GGSL discrepancy.

\bibliography{bibliography}{}
\bibliographystyle{aasjournal}

\end{document}